\documentclass[12pt]{article} %fullpage,doublespace
\usepackage{amsmath}
\usepackage{amsfonts}
\usepackage[utf8]{inputenc}
\usepackage{amssymb}
\usepackage{graphicx}
\usepackage[english]{babel}
\usepackage[T1]{fontenc} 
\usepackage{ragged2e}
\usepackage[tmargin=1.5in, bmargin=1.5in, lmargin=1.0in, rmargin=1.0in]{geometry}
\usepackage{multirow}
\usepackage{natbib}
\usepackage{color}
\usepackage[table]{xcolor}
\usepackage{colortbl}
\usepackage{rotating}
\usepackage{setspace}
\usepackage{url}
\usepackage{epstopdf} 
\usepackage{epsfig}
\usepackage{pdflscape}
\usepackage{float}

\def\etal{~{\it et al.}}

\def\P{{\mathbb P}}
\def\E{{\mathbb E}}
\def\erf{\textrm{erf}}

\def\L{{\mathcal L}}
\def\F{{\mathcal F}}

\title{The Reactive Beta Model }

\author{
Sebastien Valeyre
  \footnote{John Locke Investments, 38 Avenue Franklin Roosevelt, 77210 Fontainebleau-Avon, France, and
  Universit\'e de Paris XIII, Sorbonne Paris Cit\'e, 93430 Villetaneuse, France}  \and
Denis Grebenkov
  \footnote{Laboratoire de Physique de la Mati\`{e}re Condens\'{e}e (UMR 7643), CNRS -- Ecole Polytechnique, 
University Paris-Saclay, 91128 Palaiseau, France} \and
Sofiane Aboura
  \footnote{Universit\'e de Paris XIII, Sorbonne Paris Cit\'e, 93430 Villetaneuse, France}} 

\date{\today}

\begin{document}
\maketitle

\begin{abstract}
\justifying 
\noindent
We present a reactive beta model that includes the leverage effect to
allow hedge fund managers to target a near-zero beta for market
neutral strategies.
For this purpose, we derive a metric of correlation with leverage
effect to identify the relation between the market beta and volatility
changes.
An empirical test based on the most popular market neutral strategies
is run from 2000 to 2015 with exhaustive data sets including 600 US
stocks and 600 European stocks.
Our findings confirm the ability of the reactive beta model to
withdraw an important part of the bias from the beta estimation
and from most popular market neutral strategies.
\\


\end{abstract}

\pagebreak

\section{Introduction}

%We propose a new measure of market beta that serves for portfolio
%hedging in market neutral strategies.  
The correct measurement of market betas is paramount for market
neutral hedge fund managers who target a near-zero beta.  Contrary to
common belief, perfect beta neutral strategies are difficult to
achieve in practice, as the mortgage crisis in 2008 exemplified, when
most market neutral funds remained correlated with stock markets and
experienced considerable unexpected losses.
This exposure to the stock index \citep{Banz81, Fama92, Fama93,
Carhart97, Ang06} is even stronger during down market conditions
\citep{Mitchell01, Agarwal04, Bussiere15}.  In such a period of market
stress, hedge funds may even add no value \citep{Asness01}.
%
%To address this complex issue, we first introduce the most popular
%beta neutral strategies, then we explain the nature of the problem
%with their practical implementation, and finally, we show how to
%resolve this problem.
 
In this paper, we test the quality of hedging for  four popular
strategies that have often been used by hedge funds.  The first and
most important strategy captures the low beta anomaly
\citep{Black72,Black72b,Haugen75,Haugen91,Ang06,Baker13,Frazzini14,Hong16}
that defies the conventional wisdom on the risk and reward trade-off
predicted by the CAPM \citep{Sharpe64}.  According to this anomaly,
high beta stocks underperform low beta stocks.  Similarly, stocks with
high idiosyncratic volatility earn lower returns than stocks with low
idiosyncratic volatility \citep{Malkiel97,Goyal03,Ang06,Ang09}. 
The related strategy consists in shorting high beta stocks and buying
low beta stocks.
The second important strategy captures the size effect
\citep{Banz81,Reinganum81,Fama92}, in which stocks of small firms tend to
earn higher returns, on average, than stocks of larger firms.  The
related strategy consists in buying stocks with small market
capitalization and shorting those with high market capitalization.
The third strategy captures the momentum effect
\citep{Jegadeesh93,Carhart97,Grinblatt04,Fama12}, where past winners
tend to continue to show high performance.  This strategy consists in
buying the past year's winning stocks and shorting the past year's
losing ones.
The forth strategy captures the short-term reversal effect
\citep{Jegadeesh90}, where past winners on the last month
tend to show low performance.  This strategy consists in buying the
past month's losing stocks and shorting the past month's winner ones
and would be highly profitable if there were no transaction cost and
no market impact.
Testing the quality of the hedge of the strategies is equivalent
to assess the quality of the beta measurements that is difficult to
realize directly as the true beta is not known.
 
The implementation of all these strategies requires a reliable
estimation of the betas to maintain the hedge.  The Ordinary Least
Squares (OLS) estimation remains the most frequently employed method,
even though it is impaired in the presence of outliers, notably from
small companies \citep{Fama08}, illiquid companies
\citep{Amihud02,Acharyaa05,Ang13}, and business cycles
\citep{Ferson99}.  In these circumstances, the OLS beta estimator
might be inconsistent.
%
%While more sophisticated regressions were proposed, 
To overcome these limitations, our approach consists in
renormalizing the returns to make them closer to Gaussian and thus to
make the OLS estimator more consistent.
In addition, many papers report that betas are time varying
\citep{Blume71,Fabozzi78,Jagannathan96,Fama97,Bollerslev88,Lettau01,Lewellen06,Ang07,Engle16}.
This can lead to measurement errors that could create serious bias in
the cross-sectional asset pricing test
\citep{Shanken92,Chan92,Meng11,Bali17}.
In fact, firms' stock betas do change over time for several reasons.
The firm's assets tend to vary over time via acquiring or replacing
new businesses that makes them more diversified.  The betas also
change for firms that change in dimension to be safer or riskier.  For
instance, financial leverage may increase when firms become larger, as
they can issue more debt.  Moreover, firms with higher leverage are
exposed to a more unstable beta \citep{Galai76,DeJong85}.
One way to account for the time dependence of beta is to consider
regime changes when the return history used in beta estimation is long
enough.  Surprisingly, only one paper \citep{Chen05} suggests a
solution to capture the time dependence and discusses regime changes
for the beta using a multiple structural change methodology.  The
study shows that the risk related to beta regime changes is rewarded
by higher returns.
Another way is to examine the correlation dynamics.
\citet{Francis79}  finds that ``the correlation with the market is
the primary cause of changing betas ... the standard deviations of
individual assets are fairly stable''. 
This finding calls for special attention to the correlation dynamics
addressed in our paper but apparently insufficiently investigated in
other works.
 
Despite the extended literature on this issue, little attention has
been paid to the link between the leverage effect\footnote{ Note that
we are not dealing with the restricted definition of the ``leveraged
beta'' that comes from the degree of leverage in the firm's capital
structure.} and the beta.  The leverage effect is defined as the
negative correlation between the securities' returns and their
volatility changes.  This correlation induces residual correlations
between the stock overperformances and beta changes.  In fact, earlier
studies have heavily focused on the role of the leverage effect on
volatility
\citep{Black76,Christie82,Campbell92,Bekaert00,Bouchaud01,Valeyre13}.
Surprisingly, despite its theoretical and empirical underpinnings, the
leverage effect has not been considered so far in beta modeling, while
it is a measure of risk.  We aim to close this gap.  Our paper starts
by investigating the role of the leverage effect in the correlation
measure by extending the reactive volatility model
\citep{Valeyre13}, which efficiently tracks the implied volatility by
capturing both the retarded effect induced by the specific risk and
the panic effect, which occurs whenever the systematic risk becomes
the dominant factor.
This allows us to set up a reactive beta model incorporating three
independent components, all of them contributing to the reduction of
the bias of the hedging.  First, we take into account the leverage
effect on beta, where the beta of underperforming stocks tends to
increase.  Second, we consider a leverage effect on correlation, in
which a stock index decline induces an increase in correlations.
Third, we model the relation between the relative volatility (defined
as the ratio of the stock's volatility to the index's volatility) and
the beta.  When the relative volatility increases, the beta increases
as well.  All three independent components contribute to the reduction
of biases in the naive regression estimation of the beta and therefore
considerably improve hedging strategies.
 
The main contribution of this paper is the formulation of a
\emph{reactive beta model with leverage effect}.  The economic
intuition behind the reactive beta model is the derivation of a
suitable beta measure allowing the implementation of the popular
market neutral hedging strategies with reduced bias and smaller
standard deviation.  In contrast, portfolio managers who use naive
beta measures remain exposed to systematic risk factors that create
biases in their market neutral strategies.
An empirical test is performed based on an exhaustive dataset that
includes 600 American stocks and 600 European stocks from the S\&P
500, Nasdaq 100, and Euro Stoxx 600 over the period from 2000 to 2015,
which includes several business cycles.  This test validates the
superiority of the reactive beta model over conventional methods.
 
The article is organized as follows.  Section \ref{sec:model} outlines
the methodology employed for the reactive beta model.  Section
\ref{sec:empirical} describes the data and empirical findings.  
Section \ref{sec:robustness} provides several robustness checks to
assess the quality of the reactive beta model against alternative
methods.  Section \ref{sec:application} expands the discussion beyond
the field of portfolio management, while Section
\ref{sec:conclusion} concludes.

\section{The reactive beta model}
\label{sec:model}

In this section, we present the reactive beta model with three
independent components.  First, we take into account the specific
leverage effect on beta.  Second, we consider the systematic leverage
effect on correlation.  Third, we model the relation between the
relative volatility and the beta via the nonlinear beta
elasticity.

\subsection{The leverage effect on beta}
\label{sec:RVM}

We first account for relations between returns, volatilities, and
beta, which are characterized by the so-called leverage effect.  This
component takes into account the phenomenon when a beta increases as
soon as a stock underperforms the index.  Such a phenomenon can be
fairly well described by the leverage effect captured in the reactive
volatility model.  We call the \textit{specific leverage effect} the
negative relation between specific returns and the risk (here, the
beta), where the specific return is the non-systematic part of the
returns (a stock's overperformance).  The specific leverage effect on
beta follows the same dynamics as the specific leverage effect
introduced in the reactive volatility model.

\subsubsection{The reactive volatility model}
\label{sec:stock_index}

This section aims at capturing the dependence of betas on stock
overperformance (when a stock is overperforming, its beta tends to
decrease).  For this purpose, we rely on the methodology of the
reactive volatility model \citep{Valeyre13} to derive a stable measure
of beta by using the renormalization factor that depends on the
stock's overperformance.  The model describes the systematic and
specific leverage effects.  The systematic leverage due to the panic
effect and the specific leverage due to a retarded effect have very
different time scales and intensity.  These two different effects were
investigated by \citet{Bouchaud01,Valeyre13}.

We start by recalling the construction of the reactive volatility
model, which explicitly accounts for the leverage effect on
volatility.  Let $I(t)$ be a stock index at day $t$.  It is well known
that arithmetic returns, $r_I(t) = \delta I(t)/I(t-1)$, are
heteroscedastic, partly due to price-volatility correlations.
Throughout the text, $\delta$ refers to a difference between
successive values, e.g., $\delta I(t) = I(t)-I(t-1)$.  The reactive
volatility model aims at constructing an appropriate ``level'' of the
stock index, $L(t)$, to substitute the original returns $\delta
I(t)/I(t-1)$ by less-heteroscedastic returns $\delta I(t)/L(t-1)$.

For this purpose, we first introduce two ``levels'' of the stock index
as exponential moving averages (EMAs) with two time scales: a slow
level $L_s(t)$ and a fast level $L_f(t)$.  In addition, we denote by
$L_{is}(t)$ the EMA (with the slow time scale) of the price $S_i(t)$
of the stock $i$ at time $t$.  These EMAs can be computed using
standard linear relations:
\begin{eqnarray}
\label{eq:Lslow}
L_s(t) &=& (1-\lambda_s) L_s(t-1) + \lambda_s I(t) , \\
\label{eq:Lf}
L_f(t) &=& (1-\lambda_f) L_f(t-1) + \lambda_f I(t) , \\
L_{is}(t) &=& (1-\lambda_s) L_{is}(t-1) + \lambda_s S_{i}(t) ,
\end{eqnarray}
where $\lambda_s$ and $\lambda_f$ are the weighting parameters of the
EMAs that we set to $\lambda_s = 0.0241$ and $\lambda_f = 0.1484$,
relying on the estimates by \citet{Bouchaud01}.  The slow parameter
corresponds to the relaxation time of the retarded effect for the
specific risk, whereas the fast one corresponds to the relaxation time
of the panic effect for the systematic risk.  These two relaxation
times are found to be rather universal, as they are stable over the
years and do not change among different mature stock markets.
The appropriate levels $L(t)$ and $L_i(t)$, accounting for the
leverage effect on the volatility, were introduced for the stock index
and individual stocks, respectively
\footnote{
In practice, a filtering function is introduced to attenuate the
contribution from eventual extreme events or wrong data (outliers).
The filter was applied to $z=\frac{L_s(t)-I(t)}{I(t)}$ and
$z=\frac{L_{is}(t)-S_i(t)}{S_i(t)}$ in Eqs. (\ref{eq:L_Taylor},
\ref{eq:Li_Taylor}) and was defined as $F_\phi(z) = \tanh(\phi
z)/\phi$ with $\phi = 3.3$ (in the limit $\phi = 0$, there is no
filter: $F_0(z) = z$).  }
%footnote
%
\begin{eqnarray}
\label{eq:L_Taylor}
L(t) & = & I(t) \left(1 + \frac{L_s(t)-I(t)}{I(t)}\right) \left(1 + \ell ~ \frac{L_f(t) - I(t)}{L_f(t)}\right) , \\
\label{eq:Li_Taylor}
L_{i}(t) & = &  S_i(t) \underbrace{\left(1 + \frac{L_{is}(t)-S_i(t)}{S_i(t)}\right)}_{\textrm{specific risk}}
\underbrace{\left(1 + \ell_i ~ \frac{L_f(t) - I(t)}{L_f(t)}\right)}_{\textrm{systematic risk}},
\end{eqnarray}
with the parameters $\ell$ and $\ell_i$ quantifying the leverage.
The parameter $\ell$ was defined by \citet{Valeyre13} and
deduced to be around 8 from another parameter estimated by
\citet{Bouchaud01} on 7 major stock indexes.  If $\ell =
\ell_i$, the correlation between the stock index and the individual
stock $i$ is not impacted by the leverage effect.  In turn, if $\ell >
\ell_i$, the correlation increases when the stock index decreases.
Although $\ell_i$ can generally be specific to the considered $i$-th
stock, we ignore its possible dependence on $i$ and set $\ell_i =
\ell'$.
Using the levels $L(t)$ and $L_i(t)$, we introduce the normalized
returns:
\begin{equation}  \label{eq:tilder}
\tilde r_I = \tilde r_I(t) = \frac{\delta I(t)}{L(t-1)}, \qquad
\tilde r_i = \tilde r_i(t) = \frac{\delta S_i(t)}{L_i(t-1)}
\end{equation}
and compute the renormalized variances $\tilde{\sigma}_I^2$ and
$\tilde{\sigma}_i^2$ through the EMAs as:
\begin{eqnarray}
\label{eq:sigmaIt0}
\tilde{\sigma}_I^2(t) &=& (1 - \lambda_\sigma) \tilde{\sigma}_I^2(t-1) + \lambda_\sigma \tilde r_I^2(t) ,   \\
\tilde{\sigma}_i^2(t) &=& (1 - \lambda_\sigma) \tilde{\sigma}_i^2(t-1) + \lambda_\sigma \tilde r_i^2(t) ,
\end{eqnarray}
where $\lambda_\sigma$ is a weighting parameter that has to be chosen
as a compromise between the accuracy of the estimated renormalized
volatility and the reactivity of that estimation.  Indeed, the
renormalized returns are constructed to be homoscedastic only at short
times because the renormalization based on the leverage effect with
short relaxation times ($\lambda_s$, $\lambda_f$) cannot account for
long periods of changing volatility related to economic cycles.  Since
economic uncertainty does not change significantly in a period of two
months (40 trading days), we set $\lambda_\sigma$ to $1/40 = 0.025$.
This sample length leads to a statistical uncertainty of approximately
$\sqrt{1/40} \approx 16\%$.  Finally, these renormalized variances can
be converted into the reactive volatility $\sigma_I(t)$ of the stock
index quantifying the systematic risk governed by the panic effect,
and the reactive volatility $\sigma_i(t)$ of each individual stock
quantifying the specific risk governed by the leverage effect:
\begin{eqnarray}
\label{eq:sigmaInew}
\sigma_{I}(t) &=& \tilde{\sigma}_{I}(t) \frac{L(t)}{I(t)} , \\
\label{eq:sigmainew}
\sigma_{i}(t) &=& \tilde{\sigma}_{i}(t) \frac{L_i(t)}{S_i(t)} .
\end{eqnarray}

This reactive volatility captures a large part of the
heteroscedascticity, i.e., a large part of the volatility variation is
completely explained by the leverage effect.  For instance, if the
stock index loses 1\%, $\frac{L(t)}{I(t)}$ increases by $\ell \times
1\%=8\%$, and stock index volatility increases by 8\%.  That is enough
to capture the large part of the VIX variation, with $R^2 = 0.46$, see
Fig. 4 by \citet{Valeyre13}.  In turn, if the stock underperforms the
stock index by 1\%, $\frac{L_i(t)}{S_i(t)}$ increases by 1\%, and the
single stock volatility increases by 1\%.

\subsubsection{The specific leverage effect on beta}
\label{sec:beta_specific}

The volatility estimation procedure naturally impacts the estimation
of beta.  Many financial instruments rely on the estimated beta,
$\beta_i$, which corresponds to the slope of a linear regression of
stocks' arithmetic returns $r_i$ on the index arithmetic return $r_I$:
\begin{equation}
\label{eq:simplemarketmodel}
r_i = \beta_i r_I + \epsilon_i,  \qquad \textrm{with} \quad 
r_i = \frac{\delta S_i(t)}{S_i(t-1)}, \quad   r_I = \frac{\delta I(t)}{I(t-1)} ,
\end{equation}
where $\epsilon_i$ is the residual random component specific to 
stock $i$.  We consider another beta estimate, $\tilde \beta_i$, based
on the reactive volatility model, in which the renormalized stock
returns $\tilde r_i$ are regressed on the renormalized stock index
returns $\tilde r_I$:
\begin{equation}
\label{eq:simplereactivemodel}
\tilde r_{i} = \tilde\beta_i \, \tilde r_{I} + \tilde \epsilon_{i}, \qquad \textrm{with} \quad
\tilde r_{i} = \frac{\delta S_{i}(t)}{L_{i}(t-1)}, \quad  \tilde r_{I} = \frac{\delta I(t)}{L(t-1)}.
\end{equation}
We then obtain a reactive beta measure:
\begin{equation}
\label{eq:reactivevolatility}
\beta_{i}(t) = \tilde \beta_{i}(t) \frac{\sigma_{i}(t) \, \tilde \sigma_{I}(t)}{\sigma_{I}(t)\, \tilde \sigma_{i}(t)} 
=\tilde \beta_{i}\frac{L_{is}(t) I(t)}{L_{s}(t) S_{i}(t)} ,
\end{equation}
which includes two improvements:
\begin{itemize}
\item 
$\tilde \beta_{i}$, which becomes less sensitive to price changes by
accounting for the specific leverage effect;

\item  
$\sigma_{i} \tilde \sigma_{I}/(\sigma_{I} \tilde \sigma_{i})$, which
changes instantaneously with price changes.
\end{itemize}

When taking into account the short-term leverage effect in
correlations, the reactive term is reduced to $\frac{L_{is}(t)
I(t)}{L_{s}(t) S_{i}(t)}$.  This term has a significant impact, as
the beta of underperforming stocks should increase.

\subsection{The systematic leverage effect on correlation}
\label{sec:RCM}

\subsubsection{The empirical estimation of $\ell'$}

\begin{figure}
\begin{center}
\includegraphics[width=120mm]{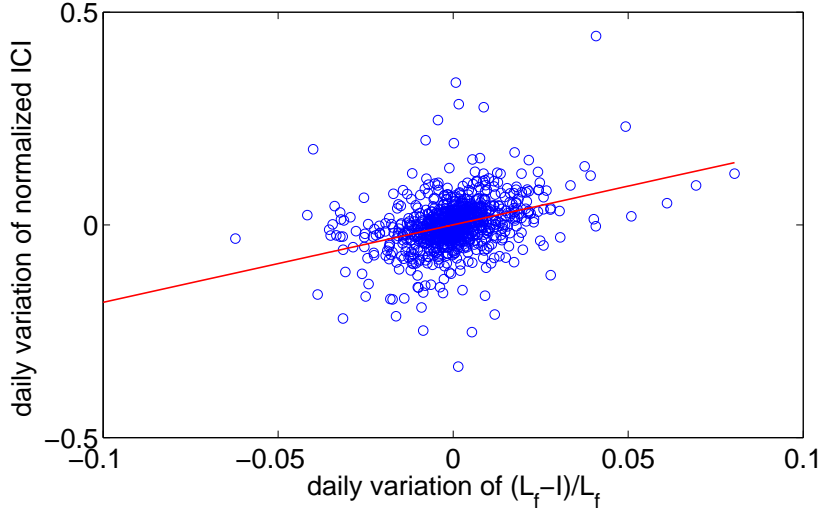}
\end{center}
\caption{
Daily variations of the CBOE S\&P 500 Implied Correlation Indices
(ICI) since their inception, divided by their mean, versus daily
variations of the leverage factor $(L_f-I)/L_f$.  A linear regression
(solid line) yields the coefficient $1.82 \pm 0.16$ (i.e.,
$2(\ell-\ell^\prime) = 1.82$), with $R^2 = 0.13$ and t-statistics of
$11.4$. Period: 2007-2015.}
\label{fig:RegVarRho}
% JLI_Seb_RCM_fig1();
\end{figure}

We coin by \textit{systematic leverage effect} the negative relation
between systematic returns and the risk (here, the correlation), where
the systematic returns are the non-specific part of the returns (stock
index performance).  The systematic leverage effect on correlation
follows the same dynamics as the systematic leverage effect introduced
in the reactive volatility model (the phenomenon's duration is
approximately 7 days for $\lambda_f=0.1484$).
All correlations are impacted together in the same way by the
systematic leverage effect, and single stocks and their stock indexes
should also shift in the same direction.  This explains why the
stock's beta will not change with respect to the index.  The
implication is that betas are not very sensitive to the systematic
leverage effect, in contrast to the specific leverage effect.
We consider the impact of the short-term systematic leverage effect on
correlation.  Assuming that the correlation between each individual
stock and the stock index is the same for all stocks, one can define
the implied correlation as:
\footnote{
See \url{http://www.cboe.com/micro/impliedcorrelation/impliedcorrelationindicator.pdf}}
\begin{equation}
\label{eq:rho_Taylor}
\rho(t) = \frac{\sigma^{2}_{I}(t) - \sum\limits_{i}{w_{i}^{2}\sigma_{i}^{2}(t)}}{\sum\limits_{i\neq j} w_{i}w_{j}\sigma_{i}(t)\sigma_{j}(t)} ,
\end{equation}
where $w_i$ represents the weight of stock $i$ in the index.
Denoting
\begin{equation}
\label{eq:e_Taylor}
e_{I}(t) = \frac{\hat{L}_s(t)}{I(t)}  - 1 , \qquad
e_{i}(t) = \frac{\hat{L}_{is}(t)}{S_i(t)} - 1 ,
\end{equation}
we use Eqs. (\ref{eq:sigmaInew}, \ref{eq:sigmainew}) to obtain:
\begin{equation}
\label{eq:rho_Taylor2}
\rho = \frac{\tilde{\sigma}^2_I (1+e_{I})^2 \left(1+ \ell\frac{L_f - I}{L_f} \right)^2
- \left(1+ \ell^{\prime}\frac{L_f - I}{L_f} \right)^2 \sum\limits_{i} w_i^2 (1+e_i)^2 \sigma_i^2 }
{\left(1+ \ell^{\prime} \frac{L_f - I}{L_f} \right)^2 \sum\limits_{i\neq j} w_i w_j \tilde{\sigma}_i\tilde{\sigma}_j(1+e_i)(1+e_j)} .
\end{equation}
If the weights $w_i$ are small, we can ignore the second term; in
addition, if $e_i$ are small, then
\begin{equation*}
\sum\limits_{i\neq j} w_i w_j \tilde{\sigma}_i \tilde{\sigma}_j (1+e_i)(1+e_j) \approx (1+e_I)^2 \tilde\sigma_0^2 ,
\end{equation*}
where $\tilde\sigma_0^2$ is an average of $\tilde\sigma_i^2$.  Keeping
only the leading terms of the expansion in terms of the small
parameter $(L_f-I)/L_f$, one thus obtains
\begin{equation}
\label{eq:modelcorrel}
\rho \approx \frac{\tilde{\sigma_{I}}^2}{\tilde{\sigma_{0}}^2}\left( 1+2(\ell-\ell^{\prime})\frac{L_{f} - I}{L_{f}}\right) .
\end{equation}
This relation shows the dynamics of the implied correlation $\rho$
induced by the leverage effect (accounted through the factor $(L_f
-I)/L_f$).  We assume that the same dynamics are applicable to
correlations between individual stocks, i.e.,
\begin{equation}
\label{eq:correl}
\rho_{i,j} = \tilde \rho_{i,j}\left( 1+2(\ell-\ell^{\prime})\frac{L_{f} - I}{L_{f}}\right),
\end{equation}
where $\tilde{\rho}_{i,j}$ are the parameters specific to each pair of
stocks $i$ and $j$.  From this relation, we derive a measure of
correlation accounting for the leverage effect between the single
stock $i$ and the stock index:
\begin{equation}
\label{eq:correli}
\rho_i = \tilde \rho_i\left( 1+(\ell-\ell^{\prime})\frac{L_{f} - I}{L_{f}}\right) ,
\end{equation}
where $\tilde{\rho}_i$ are the parameters specific to each stock $i$.
Note that there is no factor 2 in front of $(\ell-\ell^\prime)$ in
Eq. (\ref{eq:correli}) because we have a one-factor model here.  We
use Eq. (\ref{eq:correli}) in the reactive beta model (see
Eqs. (\ref{eq:eq1}, \ref{eq:beta_final}) below) to take into account
the varying nature of the correlation in the regression.  We rescale
the measurement by the normalization factor $(1+(\ell -\ell^\prime)
(L_f-I)/L_f)$ and then recover the variation of the correlation
through the denormalization factor
$1/(1+(\ell-\ell^\prime)(L_f-I)/L_f)$.
We emphasize that the parameter $\ell$ in Eq. (\ref{eq:L_Taylor}) that
quantifies the systematic leverage for the stock index is slightly
different from the parameter $\ell'$ in Eq. (\ref{eq:Li_Taylor}) that
quantifies the systematic leverage for single stocks.
According to Eq. (\ref{eq:correl}), when the market decreases,
correlations between stocks increase as $\ell > \ell^\prime$, and
therefore, the stock index volatility increases more than the single
stocks volatility: $\delta (\sigma_i/\sigma_I)<0$.  Once again, the
beta is, in contrast to the correlation, weakly impacted by the
systematic leverage effect, as all correlations increase in the same
way.  More precisely, it means that the impact of the increase of
correlation in the beta measurement is compensated by a decrease of
the relative volatility: $\delta (\sigma_i/\sigma_I)<0$, i.e., the
single stock volatility increase is lower than that of the stock index
volatility.  For this reason, the reactive beta model in
Eqs. (\ref{eq:eq1}, \ref{eq:beta_final}) is not very sensitive to the
choice of $\ell'$.  Nevertheless, we explain in this section how
$\ell'$ is calibrated using the implied volatility index.  We measure
the level of the systematic leverage effect $\ell^\prime$ for a single
stock by regressing Eq. (\ref{eq:modelcorrel}) with data from the
market-implied correlation S\&P 500 index.  Figure~\ref{fig:RegVarRho}
illustrates the slope of this regression.  By regressing
$\frac{L_f-I}{L_f}$ against $\frac{\rho}{\tilde{\rho}_0}$, where
$\tilde{\rho}_0$ is the average of $\rho$, we deduce that empirically
we can set:
\begin{equation} \label{eq:ell_diff}
\ell-\ell^{\prime} = 0.91\pm 0.08  ,
\end{equation}
with a t-statistics of $11.4$.
Since $\ell-\ell^{\prime} \ll \ell (= 8)$, we deduce an important
result, namely, that the systematic leverage impact on the correlation
is more than 8 times smaller than the systematic leverage impact on
volatility.  The main consequence is that although statistically
significant, the leverage effect is not a major component of the
correlation.

\subsubsection{The systematic leverage effect component in the reactive model}

As just discussed, the correlation increases when stock index
price decrease.  This effect could generate a bias in the beta
measurement as stock index prices could fluctuate in a sample used to
measure the slope.  Our solution is to adjust the beta between
renormalized returns through the correction factor $\L(t)$ defined as
\begin{equation}
\label{eq:L}
\L(t) = 1+(\ell-\ell^{\prime}) \left(\frac{L_f(t-1)-I(t-1)}{L_f(t-1)}\right),  
\end{equation}
The correction factor $\L(t)$ should be used to estimate the slope
between the stock index and single stock returns and then to
denormalize the slopefor getting the reactive beta that depends
directly on $\L(t)$.  
%From Eq. (\ref{eq:reactivevolatility}), we
%obtain therefore:
%%
%\begin{equation}
%\label{eq:reactivevolatilityL}
%\beta_{i}(t) =\tilde \beta_{i}\frac{L_{is}(t) I(t)}{L_{s}(t) S_{i}(t)} \L(t).
%\end{equation}

\subsection{The relation between the relative volatility and beta}
\label{sec:beta_relation}

\subsubsection{The empirical estimation of beta elasticity}

\begin{figure}
\begin{center}
\includegraphics[width=120mm]{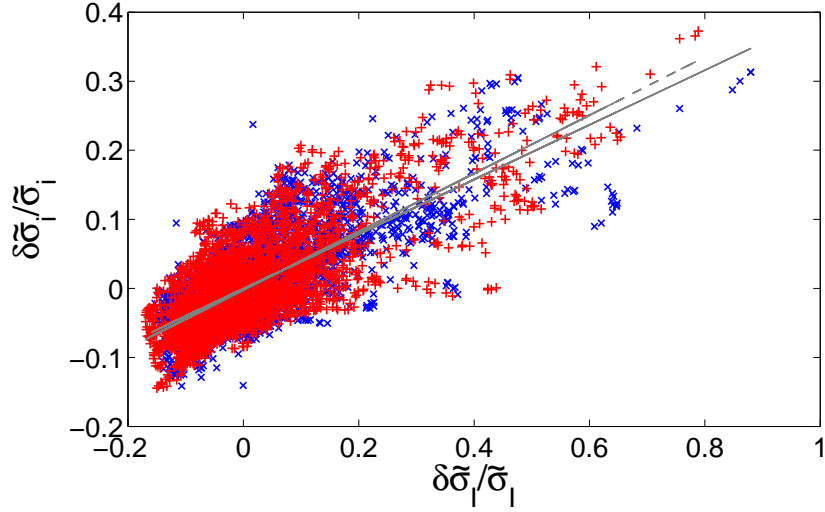}
\end{center}
\caption{
Normalized daily variations of $\tilde\sigma_{i}$, $\delta \tilde
\sigma_i/\tilde \sigma_i = \frac{\tilde
\sigma_i(t)-\tilde\sigma_i(t-1)}{\tilde \sigma_i(t-1)}$, versus
normalized daily variations of $\tilde\sigma_{I}$, $\delta
\tilde \sigma_I/\tilde \sigma_I = \frac{\tilde
\sigma_I(t)-\tilde\sigma_I(t-1)}{\tilde \sigma_I(t-1)}$,
for the European market (blue crosses) and the US market (red pluses).
The two gray lines show the linear regression of both datasets, with
regression coefficients of $0.40$ ($R^2=0.60$) and $0.42$ (with
$R^2=0.59$) for the European and US markets, respectively.  The time
frame includes observations from the technology bubble burst, the
U.S. subprime, and Euro debt crises. Period: 1998-2015.}
\label{fig:dsigmaidsigmaI}
\end{figure}

In this part, we identify correlations between the relative volatility
and beta changes.  We choose the relative volatility defined as the
ratio $\tilde\sigma_i/\tilde\sigma_I$ as an explanatory variable of
$\tilde\beta_i$, because $\tilde\beta_i$ is expected to be constant if
the ratio $\tilde\sigma_i/\tilde\sigma_I$ is constant.  However,
empirically, the ratio $\tilde\sigma_i/\tilde\sigma_I$ can change
dramatically between periods of high dispersion (i.e., when stocks
are, on average, weakly correlated) and low systematic risk (i.e.,
when stock indexes are not stressed), and periods of low dispersion
and high systematic risk.  Figure \ref{fig:dsigmaidsigmaI}
illustrates, for both European and US markets, that the dispersion
among stocks decreases, on average, when markets become volatile.  A
linear regression of rescaled daily variations of $\tilde\sigma_i$
yields:
\begin{equation}
\label{eq:delta_sigma_auxil1}
\frac{\delta \tilde\sigma_i(t)}{\tilde\sigma_i(t-1)} \approx 0.4 \,  \frac{\delta \tilde\sigma_I(t)}{\tilde\sigma_I(t-1)} + \epsilon_i ,
\end{equation}
where $\epsilon_i$ is the residual (specific) noise.  Using the
standard rules for infinitesimal increments, we find from this
regression:
\begin{equation}
\label{eq:delta_sigma_auxil2}
\delta \left(\frac{\tilde\sigma_i}{\tilde\sigma_I}\right) \simeq \frac{\delta \tilde\sigma_i}{\tilde\sigma_I} 
- \frac{\tilde\sigma_i \, \delta \tilde\sigma_I}{\tilde\sigma_I^2}
= \frac{\tilde\sigma_i}{\tilde\sigma_I} \left(\frac{\delta \tilde\sigma_i}{\tilde\sigma_i} - \frac{\delta \tilde\sigma_I}{\tilde\sigma_I} \right)
\simeq -0.6 \frac{\tilde\sigma_i}{\tilde\sigma_I} \, \frac{\delta \tilde\sigma_I}{\tilde\sigma_I} ,
\end{equation}
i.e., the relative volatility $\tilde\sigma_i/\tilde\sigma_I$ is
relatively stable but its small variations can still impact the
beta estimation.  This empirical relation shows that when there is a
volatility shock in the market, the stock index volatility increases
much faster than the average single stock volatility.

Because we want to take into account the impact of the relative
volatility change on the beta measurement, we introduce the derivative
of the beta with respect to the logarithm of the squared relative
volatility:
\begin{equation}
\label{eq:f}
f(\tilde \beta_i) = \frac{d\tilde\beta_i}{d\ln(\tilde\sigma_i^2/\tilde\sigma_I^2)} 
= \frac{d\tilde\beta_i}{d(\tilde\sigma_i/\tilde\sigma_I)} \, \frac{\tilde \sigma_i}{2\tilde\sigma_I} .
\end{equation}
We expect that $f(\tilde\beta_i)$ is positive and increasing with
$\tilde\beta_i$.  Indeed, we expect that a stock with a low beta
should have a stable beta (less sensitive to its relative volatility
increase), as the increase in this case is most likely due to a
specific risk increase.  In such a case, the sensitivity of beta to
the relative volatility is weak.  In the opposite case of a high beta,
a stock that is highly sensitive to the stock index will face a beta
decline as soon as its relative volatility decreases.
Consequently, when there is a volatility shock in the market,
$\delta(\frac{\tilde\sigma_i}{\tilde\sigma_I})$ is negative, and
therefore, the beta of stocks with high beta and high $f$ is reduced.
In turn, the stocks with low beta are less impacted because $f$ is
smaller and $\delta(\tilde\sigma_i/\tilde\sigma_I)$ is expected to be
less negative.

When the correlation of the stock with the stock index is constant, we
can use a linear model: $f(\tilde \beta_i) = \tilde\beta_i/2$.  In
fact, using the relation $\tilde \beta_i = \tilde \rho_i \frac{\tilde
\sigma_i}{\tilde \sigma_I}$ and the assumption that $\tilde\rho_i$
is constant (i.e., it does not depend on $ \frac{\tilde
\sigma_i}{\tilde \sigma_I}$), one obtains from Eq. (\ref{eq:f}) $f =
\tilde\rho_i \frac{\tilde \sigma_i}{2\tilde \sigma_I} = \tilde\beta_i/2$.
In general, however, the correlation can depend on the relative
volatility, and thus, the function $f$ may be more complicated.
To estimate $f$, one needs the renormalized beta and the relative
volatility.  For a better estimation, we aim at reducing even further
the heteroscedasticity by using an exponential moving regression of
the returns $\tilde r_i$ and $\tilde r_I$ that are renormalized by the
estimated normalized index volatility $\tilde \sigma_I$.  We denote
these renormalized returns as:
\begin{equation}
\label{eq:hatr}
\hat r_i(t) = \frac{\tilde r_i(t)}{\tilde \sigma_I(t-1)},  \qquad  \hat r_I(t) = \frac{\tilde r_I(t)}{\tilde \sigma_I(t-1)} .
\end{equation}
Computing the EMAs, 
\begin{eqnarray}
\label{eq:SigmahatI}
\hat{\phi}_{i}(t) &=& (1-\lambda_{\beta})\hat{\phi}_{i}(t-1) + \lambda_{\beta} \hat r_i(t) \, \hat r_I(t) , \\
\label{eq:hat_sigmaI}
\hat{\sigma}_{I}^2(t) &=& (1-\lambda_{\beta})\hat{\sigma}_{I}^2(t-1) + \lambda_{\beta}\, \bigl[\hat r_I(t)\bigr]^2 ,
\end{eqnarray}
with $\lambda_\beta = 1/90$, we estimate the beta as:
\begin{equation}
\label{eq:beta}  
\hat \beta_i(t) = \frac{\hat{\phi}_{i}(t)}{\hat{\sigma}_{I}^2(t)} .
\end{equation}
Here, $\hat \phi_i$ is an estimation of the covariance between stock
index returns and single stock returns that includes two
normalizations: the levels $L_i$ and $L$ from the reactive volatility
model, and $\tilde\sigma_I$ to further reduce heteroscedasticity.  We
write $\hat \beta_i$ instead of $\tilde \beta_i$ to stress this
particular way of estimating the beta.  Similarly, the hat symbol in
Eq. (\ref{eq:hat_sigmaI}) is used to distinguish
$\hat{\sigma}_{I}(t)$, computed with renormalized index returns, from
$\tilde{\sigma}_I(t)$.  In principle, the above estimate $\tilde\beta$
could be directly regressed to the ratio of earlier estimates of
$\tilde\sigma_i$ and $\tilde \sigma_I$ from Eqs. (\ref{eq:sigmaIt0}).
However, to use the normalization by $\tilde{\sigma}_I$ consistently,
we consider the ratio of these volatilities obtained in the
renormalized form, i.e., $\hat\sigma_{i}(t)/\hat\sigma_I(t)$, where
$\hat{\sigma}_{I}(t)$ is given in Eq. (\ref{eq:hat_sigmaI}), and
\begin{equation}  \label{eq:hat_sigmai}
\hat{\sigma}^2_{i}(t) = (1-\lambda_{\beta})\hat{\sigma}^2_{i}(t-1) + \lambda_{\beta} \bigl[\hat r_i(t)\bigr]^2 . 
\end{equation}

\begin{figure}
\begin{center}
\includegraphics[width=120mm]{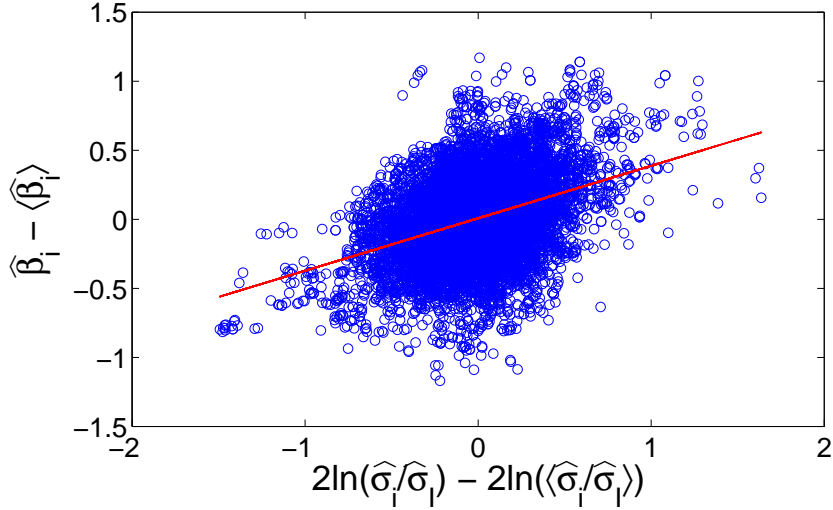}
\end{center}
\caption{
Relation between the beta $\hat{\beta}_{i}$ and the doubled logarithm
of the relative volatility $\ln(\hat{\sigma}_i/\hat{\sigma}_I)$, from
which the mean values $\langle\hat{\beta}_{i}\rangle$ and $\ln(\langle
\hat{\sigma}_i/\hat{\sigma}_I\rangle)$ were subtracted (the mean is
obtained by averaging over time for each $i$).  A linear regression is
shown by the solid line: $\hat{\beta}_{i} - \langle \hat{\beta}_{i}\rangle
= 0.76\bigl[\ln (\hat{\sigma}_i/\hat{\sigma}_I) - \ln(\langle
\hat{\sigma}_i/\hat{\sigma}_I\rangle)\bigr]$, with $R^2 = 0.14$.
For better visualization, only 10,000 randomly selected points are
shown (by circles) among 271,958 points from the European dataset.
 Period: 2014-2015.}
\label{fig:beta_log}
\end{figure}

Figure \ref{fig:beta_log} illustrates the sensitivity of beta to
relative volatilities by plotting $\hat{\beta}_{i}(t)$ from
Eq. (\ref{eq:beta}) versus $\ln(\hat{\sigma}_i(t)/\hat{\sigma}_I(t))$
for all stocks $i$ and times $t$ from 2000 to 2015, although we only
display the time frame of 2014-2015 for clarity of illustration.  On
both axes, we subtract the mean values $\langle\hat{\beta}_{i}\rangle$
and $\ln(\langle \hat{\sigma}_i/\hat{\sigma}_I\rangle)$ averaged over
all times in the whole sample.  This plot enables us to measure the
average of the $f(\tilde\beta_i)$ in Eq. (\ref{eq:f}), which is close
to $0.76/2 = 0.38$.

\begin{figure}
\begin{center}
\includegraphics[width=120mm]{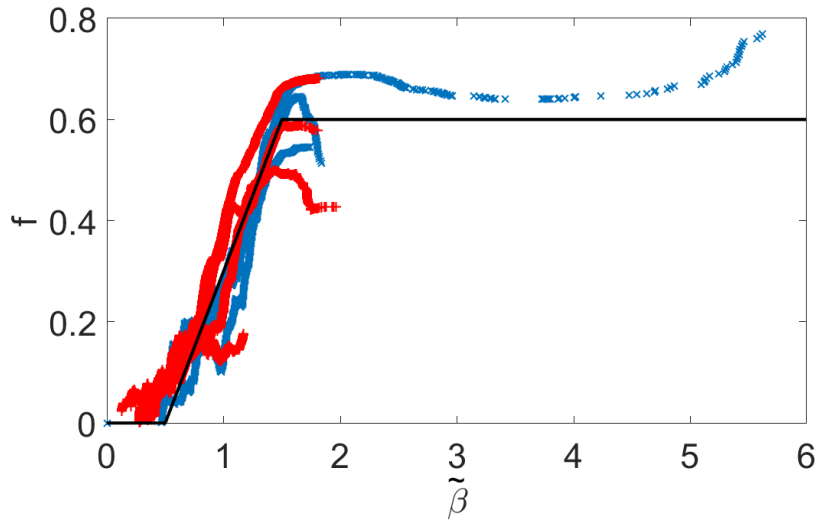}
\end{center}
\caption{
The function $f$ from Eq. (\ref{eq:f}) versus beta for the European
market (blue crosses) and the US market (red pluses).  This function
is estimated locally for 4 different time periods.  The black solid
line shows the approximation (\ref{eq:ffit}).  Period: 2000-2015.}
\label{fig:fdependingonbeta}
\end{figure}

To obtain the dependence of $f$ on beta, we estimate the slope between
$\hat{\beta}_i(t)- \langle \hat{\beta}_i\rangle$ from
Eq. (\ref{eq:beta}) and $2\ln(\hat{\sigma}_i(t)/\hat{\sigma}_I(t)) -
2\ln(\langle\hat{\sigma}_i/\hat{\sigma}_I\rangle)$ {\it locally}
around each value of $\hat \beta_i$.  For this purpose, we sort all
collected values of $\hat\beta_i$ and group them into successive
subsets, each with 10,000 points.  In each subset, we estimate the
slope between $\hat{\beta}_i(t)- \langle \hat{\beta}_i\rangle$ from
Eq. (\ref{eq:beta}) and $2\ln(\hat{\sigma}_i(t)/\hat{\sigma}_I(t)) -
2\ln(\langle\hat{\sigma}_i/\hat{\sigma}_I\rangle)$ by a standard
linear regression over 10,000 points.  This regression yields the
value of $f$ of that subset that corresponds to some average value of
$\hat\beta_i$.  Repeating this procedure over all subsets, we obtain
the dependence of $f$ on $\hat\beta_i$, which is plotted in Figure
\ref{fig:fdependingonbeta}.  We show that $f$ increases with beta.
For both European and US markets, we propose the following
approximation of the function $f$ with three different regimes:
\begin{equation}
\label{eq:ffit}
f(\tilde \beta_i) = \begin{cases} 0, \hskip 20mm \tilde\beta_{i}<0.5, \cr
0.6 (\tilde\beta_{i}-0.5) , \quad 0.5<\tilde\beta_{i}<1.6, \cr
0.6 \hskip 20mm \tilde\beta_{i}>1.6  .\end{cases}
\end{equation}
In the first regime, for low beta stocks (mostly, quality
stocks), the beta elasticity is zero that is equivalent to the
constant beta case.  For the intermediate regime, the elasticity
increases linearly with $\tilde \beta_i$ and is close to the constant
correlation case with $f(\tilde \beta_i) = \tilde \beta_i/2$.  In the
third regime for high beta stocks (speculative and growth stocks), the
elasticity is constant.  The shape of the beta elasticity is similar
for the European market and the US market.

\subsubsection{The component of the nonlinear beta elasticity}

According to Eq. (\ref{eq:ffit}), the sensitivity of the
normalized beta to changes in the relative volatility is nonlinear.
%We could interpret it as a nonlinear beta elasticity with different
%regimes.  For small beta stocks, the elasticity is zero, and beta is
%not sensitive to relative volatility variations.  For intermediate
%beta stocks, the elasticity increases linearly with beta, whereas for
%large beta stocks, the elasticity is constant.  
This elasticity could generate bias in the beta estimation if the
relative volatility changes in a sample used to measure the slope.
Our solution is to adjust the beta between normalized returns through
the correction factor $\F(t)$ defined as:
\begin{equation}
\label{eq:F}
\F(t) = 1+ \frac{2f(\tilde\beta_i(t))}{\tilde\beta_i(t)} \, \Delta \left(\frac{\tilde\sigma_i}{\tilde\sigma_I}\right) .
\end{equation}
The function $f$ is approximated by Eq. (\ref{eq:ffit}),
$\ell-\ell^\prime$ is given by Eq. (\ref{eq:ell_diff}), and
\begin{equation}
\Delta \left(\frac{\tilde\sigma_i}{\tilde\sigma_I}\right) = \frac{\tilde\sigma_i(t-1)/\tilde\sigma_I(t-1) - \sqrt{\kappa_i(t-1)}}{\sqrt{\kappa_i(t-1)}} 
\end{equation}
with
\begin{equation}
\kappa_i(t) = (1-\lambda_\beta) \kappa_i(t-1) + \lambda_\beta \biggl(\frac{\tilde\sigma_i(t)}{\tilde\sigma_I(t)} \biggr)^2
\end{equation}
being the EMA of the squared relative volatility $(\tilde
\sigma_i/\tilde \sigma_I)^2$.  The $\Delta
(\tilde\sigma_i/\tilde\sigma_I)$ quantifies deviations of the relative
volatility from its average over the sample that will be used to
estimate beta.

The correction factor $\F(t)$ should be used to estimate the slope
between stock index and single stock returns and then to denormalize
the slope for getting the reactive beta that depends directly on
$\F(t)$.  
%From Eq. (\ref{eq:reactivevolatility}) we obtain therefore:
%%
%\begin{equation}
%\label{eq:reactivevolatilityLbis}
%\beta_{i}(t) = 
%=\tilde \beta_{i}\frac{L_{is}(t) I(t)}{L_{s}(t) S_{i}(t)} \F(t)
%\end{equation}

\subsection{Summary of the reactive beta model}

In this section, we recapitulate the reactive beta model that combines
the three independent components that we described in the previous
sections: the specific leverage effect on beta, the systematic
leverage effect on correlation, and the relation between the relative
volatility and the beta.  Starting with the time series $I(t)$ and
$S_i(t)$ for the stock index and individual stocks, one computes the
levels $L_f(t)$, $L(t)$, and $L_i(t)$ from Eqs. (\ref{eq:Lf},
\ref{eq:L_Taylor}, \ref{eq:Li_Taylor}), the normalized stock index and
individual stocks returns $\tilde r_I(t)$ and $\tilde r_i(t)$ from
Eqs. (\ref{eq:tilder}), the normalized stock index volatility $\tilde
\sigma_I(t)$ from Eq. (\ref{eq:sigmaIt0}), the renormalized stock
index and individual stocks returns $\hat r_I(t)$ and $\hat r_i(t)$
from Eq. (\ref{eq:hatr}), the associated volatilities $\hat
\sigma_I(t)$ and $\hat \sigma_i(t)$ from Eqs. (\ref{eq:hat_sigmaI},
\ref{eq:hat_sigmai}), and the renormalized beta $\hat \beta_i(t)$ from
Eq. (\ref{eq:beta}).  From these quantities, one re-evaluates the
covariance between $\hat r_i$ and $\hat r_I$ by accounting for the
leverage effects and excluding the other effects.  In fact, we
compute $\hat{\Phi}_{i}(t)$ as an EMA of the normalized covariance of
the normalized daily returns:
\begin{equation}
\label{eq:eq1}
\hat{\Phi}_{i}(t) = (1-\lambda_{\beta}) \hat{\Phi}_{i}(t-1) + \lambda_{\beta} \, \frac{ \hat r_i(t) \, \hat r_I(t) }
{\L(t) \, \F(t)}  ,
\end{equation}
where $\L(t)$ and $\F(t)$ are two corrections factors defined in
Eq. (\ref{eq:L}) and Eq. (\ref{eq:F}), used to withdraw bias from the
systematic leverage and the beta elasticity.  The parameter
$\lambda_{\beta}$ describes the look-back used to estimate the slope
and is set to $1/90$ as 90 days of look-back appears to us as a good
compromise.  In fact, for a longer look-back, variations in beta,
correlation and volatilities are expected to happen due to changes of
market stress and business cycle and are not taken into account
properly by our reactive renormalization.  In turn, for a shorter
look-back, the statistical noise of the slope would be too high.

Finally, the stable estimate of the normalized beta is
\begin{equation}
\tilde\beta_i(t) = \frac{\hat{\Phi}_i(t)}{\hat\sigma_I^2(t)} ,
\end{equation}
with $\hat\sigma_I^2(t)$ defined in Eq. (\ref{eq:hat_sigmaI})
from which the estimated reactive beta of stock $i$ is deduced as
\begin{equation}
\label{eq:beta_final}
\beta_i (t) = \tilde \beta_i(t) 
\left(\frac{L_i(t) \, I(t)}{S_i(t) \, L(t)} \right)  \, \L(t) \, \F(t) .
\end{equation}
This estimation is close to the slope estimated by an OLS but with
exponentially decaying weights to accentuate recent returns and with
normalized returns to withdraw different biases.  In fact, the
normalized stable beta $\tilde \beta_i(t)$ is ``denormalized'' by the
factor that combines the three main components: the specific leverage
effect on beta, $(L_i/S_i)(I/L)$, the systematic leverage effect,
$\L(t)$, and nonlinear beta elasticity, $\F(t)$.

Every term impacts the hedging of a certain strategy:
\begin{itemize}
\item	
the term with $\L(t)$ does not have significant impact on beta, as it
is compensated in $L_i/L$, which models the short-term systematic
leverage effect on correlation in Eqs. (\ref{eq:eq1},
\ref{eq:beta_final}) (introduced in Sec. \ref{sec:RCM}), whereas the
levels $L_i$ and $L$ were introduced in the reactive volatility model.
However, it could impact the correlation by $+10\%$ if the market
decreases by $10\%$.

\item	
the term with $L_i I/(L S_i)$ that models the specific leverage effect
on volatilities (introduced in Sec. \ref{sec:beta_specific}) could
impact beta by $10\%$ if the stocks underperform by $10\%$.  This term
impacts the hedging of the short-term reversal strategy.

\item	
the term with $\F(t)$ that models the nonlinear beta elasticity which
is the sensitivity of beta to the relative volatility (introduced in
Sec. \ref{sec:beta_relation}) could impact the beta by $10\%$ if the
relative volatility increases by $10\%$.  This term impacts the
hedging of the low volatility strategy.

\end{itemize}

The simple version of the reactive beta model, when only the
leverage effect is introduced without beta elasticity and stochastic
normalized volatilities, defines an interesting class of stochastic
processes that appears to be a mean reverting with a standard
deviation linked to $\tilde{\sigma_i} \sqrt{ 1/\lambda_s}$ and a
relaxation time linked to $1/\lambda_s$.  

%{\clr Demonstration is provided in Appendix ???.}
	
The reactive beta model is based on the fit of several well identified
effects. Implied parameters work universally for all stock markets
($\ell-\ell^{\prime}$ is the only one that was fitted only on the US
market as the implied correlations for other countries are not
traded).  Here we summarize the different parameters used in the
reactive beta model:
\begin{itemize}
\item 
$\lambda_f=0.1484$ that describes the relaxation time of 7 days for
the panic effect;

\item 
$\lambda_s = 0.0241$ that describes the relaxation time of 40 days for
the retarded effect;
% , was estimated from 7 main stock indexes \citep{Bouchaud01};

\item 
$l= 8$ that describes the leverage intensity of the panic effect;
%, was estimated indirectly from 7 main stock indexes \citep{Bouchaud01};

\item 
$\ell-\ell^{\prime}\approx 0.91$ based on implied correlations on the
US stock market;

\item 
the different thresholds in the function $f(\tilde \beta_i)$ from
Eq. (\ref{eq:ffit}) that describes the nonlinear beta elasticity.
% and
%contains three different regimes.  For stocks, whose beta is lower the
%0.5, the elasticity is 0.  For stocks, whose beta is between 0.5 and
%1.6, the elasticity is linearly increasing with beta.  For stocks,
%whose beta are larger than 1.6, the elasticity is constant.  
\end{itemize}

\section{Empirical findings}
\label{sec:empirical}

\subsection{Data description}
\label{sec:data}

For the empirical calibration of $\ell-\ell^\prime$, we chose the CBOE
S\&P 500 Implied Correlation Index (ICI), which is the first widely
disseminated market-based estimate of implied average correlation of
the stocks that comprise the S\&P 500 Index (SPX).  This index begins
in July 2009, with historical data back to 2007.  We take the
front-month correlation index data from 2007 and roll it to the next
contract until the previous one expires. We also use the daily S\&P
500 stock index.
For the empirical calibration of the other parameters of the reactive
beta model, we use the daily S\&P 500 stock index and 600 largest US
stocks from January 1, 2000, to May 31, 2015.  For the European
market, we consider the EuroStoxx50 index and the 600 largest European
stocks over the same period.  The same data are used for both
calibration parameters and empirical tests.

To be precise we kept the parameters of the reactive volatility
models, that describes the intensity, the relaxation time of the
specific and systematic leverage effect that appear the most
important, identical to those that were calibrated in a period prior
to 2000 by \citep{Bouchaud01}.

\subsection{Empirical results}
\label{sec:results}

In this section, we show that exposure to the common risk factors can
sometimes lead to a high exposure of market neutral funds to the stock
market index if the betas are not correctly assessed.
Indeed, although market neutral funds should be orthogonal to
traditional asset classes, such is not always the case during extreme
moves \citep{Fung97}.  For instance, \citet{Patton09} tests the zero
correlation against non-zero correlation and finds that approximately
25\% of the market neutral funds exhibit some significant
non-neutrality, concluding that ``many market neutral hedge funds are
in fact not market neutral, but overall they are, at least, more
market neutral than other categories of hedge funds.''
The reactive beta model can help hedge funds be more market neutral
than others.  To demonstrate this, we empirically test the efficiency
of our methodology in estimating the reactive beta model using the
most popular market neutral strategies (low volatility, momentum and
size):
\begin{itemize} 

\item	
low volatility (beta) strategy: buying the stocks with the highest
$30\%$ beta and shorting those with the lowest $30\%$ beta (estimated by
the standard methodology);

\item	
short-term reversal strategy: shorting the stocks with the highest
$15\%$ one-month returns and buying those with the lowest $15\%$
one-month returns;

\item	
momentum strategy: buying the stocks with the highest
$15\%$ two-year returns and shorting those with the lowest $15\%$
two-year returns;

\item	
size strategy: buying the stocks with the highest $30\%$
capitalization and shorting those with the lowest $30\%$ capitalization.

\end{itemize}

The construction of the four most popular strategies is explained in
Appendix \ref{sec:Afactors}.  For each strategy, we compare two
different methods to estimate the beta that use only the past
information to avoid look-ahead bias: the ordinary least square (OLS)
(that is equivalent to our model with $L_i = S_i$, $L = I$, $\ell =
\ell' = 0$, and $f = 0$, with the same exponential weighting scheme)
and our reactive method.  We analyze two statistics:
\begin{itemize} 
\item 
Statistics 1: the CorSTD, that describes the unrobustness of the hedge
and in consequence the inefficiency of the beta measurement.  The CorSTD
is defined as the standard deviation of the 90-day correlation of the
strategy with the stock index returns.  The more robust the strategy
is, the lower is the CorSTD statistics.  If the strategy was well hedged,
the correlation would fluctuate by approximately $0$ within the
theoretical $10\%$ standard deviation and CorSTD will be of $10\%$
($10\%$ is obtained with uncorrelated Gaussian variables for 90-day
correlations).

\item 
Statistics 2: the Bias, that describes the bias in the hedge of the
strategy and of the beta measurement, that is defined as the
correlation of the strategy with the stock index returns on the whole
period.

\end{itemize}

This statistics are a proxy for assessing the quality of the beta
measurement that is very difficult to realize directly as true beta
are not known.

Table \ref{tab:schemes} summarizes the statistics of the four
strategies for the US and Europe markets.  We see the highest bias for
the low volatility strategy when hedged with the standard approach
($-25.5\%$ for USA and $-22.4\%$ for Europe).  The CorSTD is
approximately $20\%$, i.e., twice as high as expected if the
volatility were stable, which means that the efficiency of the hedge
is time-varying.  This could represent an important risk for fund of
funds managers, where hidden risk could accumulate and arise
especially when the market is stressed.  Indeed, the bias seems to be
higher by approximately $-60\%$ for both the USA and Europe when the
market was stressed in 2008.  We see that the use of the reactive beta
model reduces the bias in the low volatility factor, and that the
residual bias comes from the selection bias (see Appendix
\ref{sec:selection}).  When using the OLS, the possible
loss in 2008 would have been $-9.6\%$ ($= -60\% \times 40\% \times
8\%/20\%$) for a $40\%$ stock decline with a fund invested entirely on
a low volatility anomaly with a bias of $-60\%$ and a target annualized
volatility $8\%$ for the fund and $20\%$ for the index. 

We also see a significant bias for the short-term reversal strategy
when hedged with the standard approach (approximately $13.1\%$ in the
USA and in Europe).  The CorSTD is approximately $18\%$.  The
efficiency of the hedge depends on the recent past performance of the
strategy.  As soon as the strategy starts to lose, the efficiency will
decline and risk will arise, as in 2009.  Again, we see that the
reactive beta model reduces the bias in the short-term reversal
factor.  The biases and CorSTD are lower for the momentum
strategy ($-6.3\%$ in the USA, with a CorSTD of $18.3\%$) and
are of same magnitude for the size strategy ($-7.6\%$ in the USA with
a CorSTD of $17.0\%$).  The reactive beta model further reduces
the bias and the CorSTD.  This is also valid for the European
market.

We conclude that the reactive beta model reduces the bias of the low
volatility factor when it is stressed by the market.  The remaining
residual is most likely explained by the selection bias (see Appendix
\ref{sec:selection} for a formal proof).  The improvement is more
significant for the momentum factors and for the size factor in the
U.S. only.

\begin{figure}%[H]
\begin{center}
\includegraphics[width=120mm]{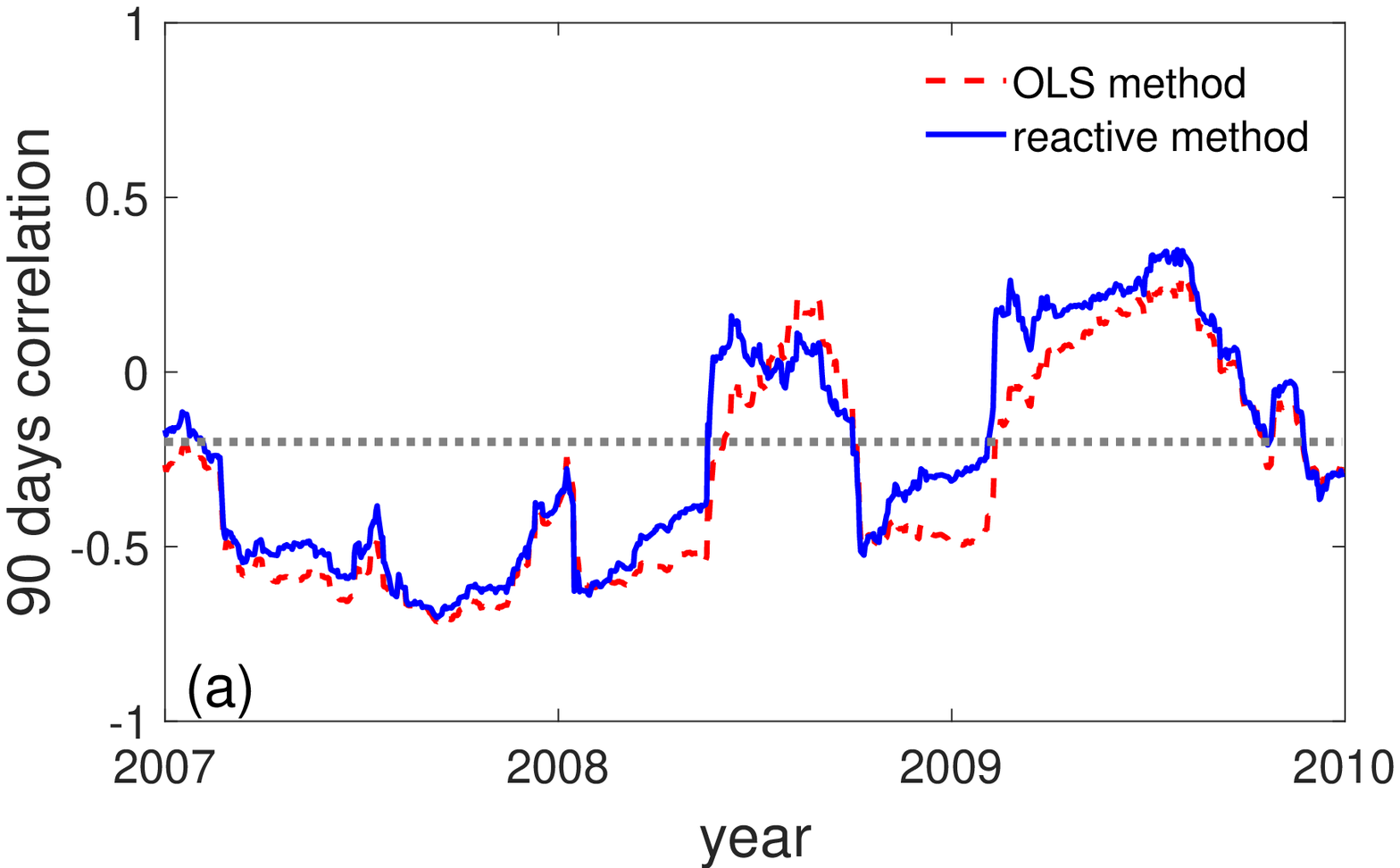}
\includegraphics[width=120mm]{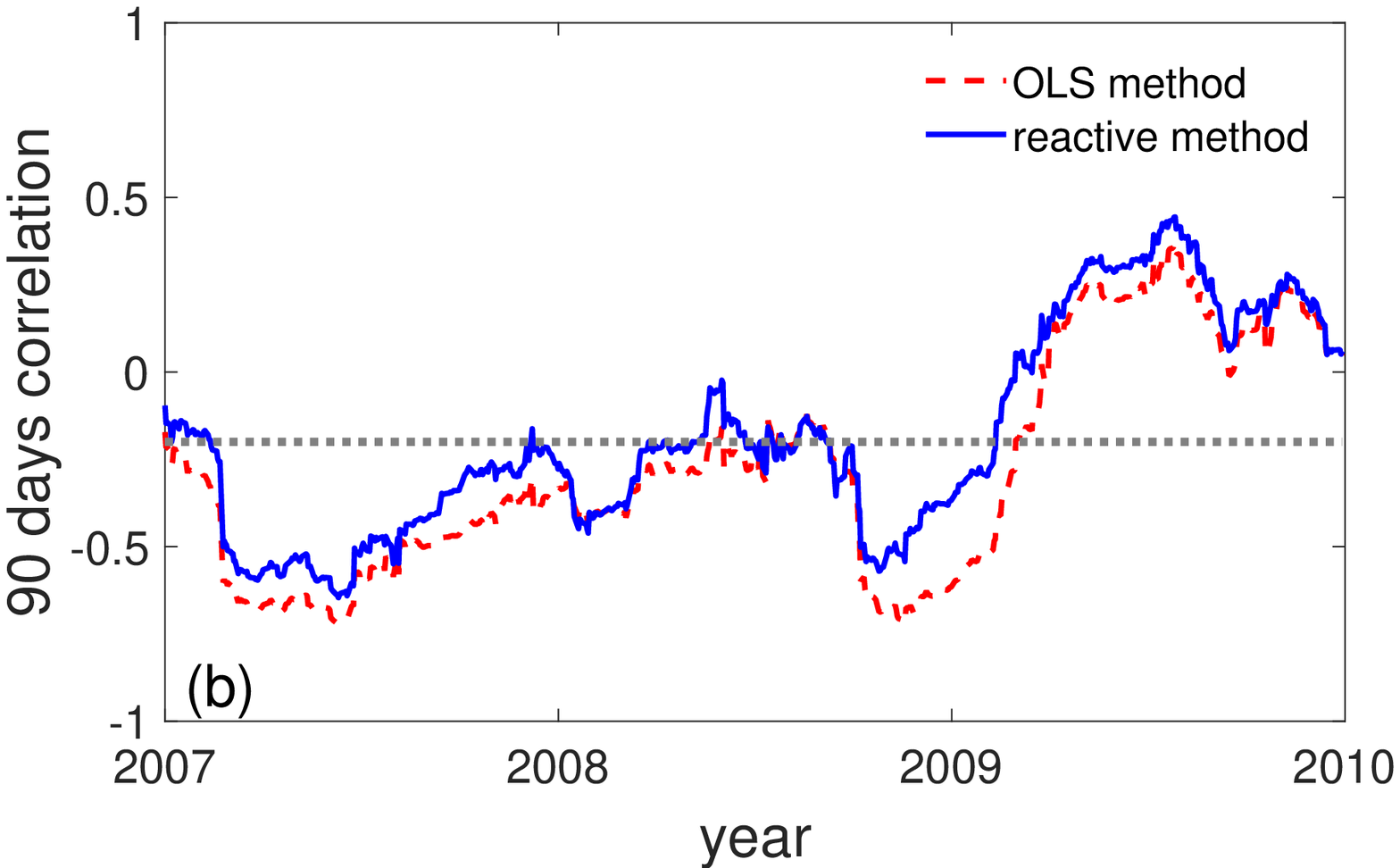}
\end{center}
\caption{
Ninety-day correlation of the low volatility factor with the stock
index in the European market {\bf (a)} and in the USA market {\bf
(b)}.  Solid and dashed lines present the proposed the reactive beta
model and the OLS methodology, respectively.  The dotted horizontal
line shows the selection bias of $-19.10\%$, as shown in Appendix
\ref{sec:selection}.  A time frame surrounding the financial crisis is
chosen.  Period: 2007-2010.}
\label{fig:betafactor}
\end{figure}

\begin{figure}%[H]
\begin{center}
\includegraphics[width=120mm]{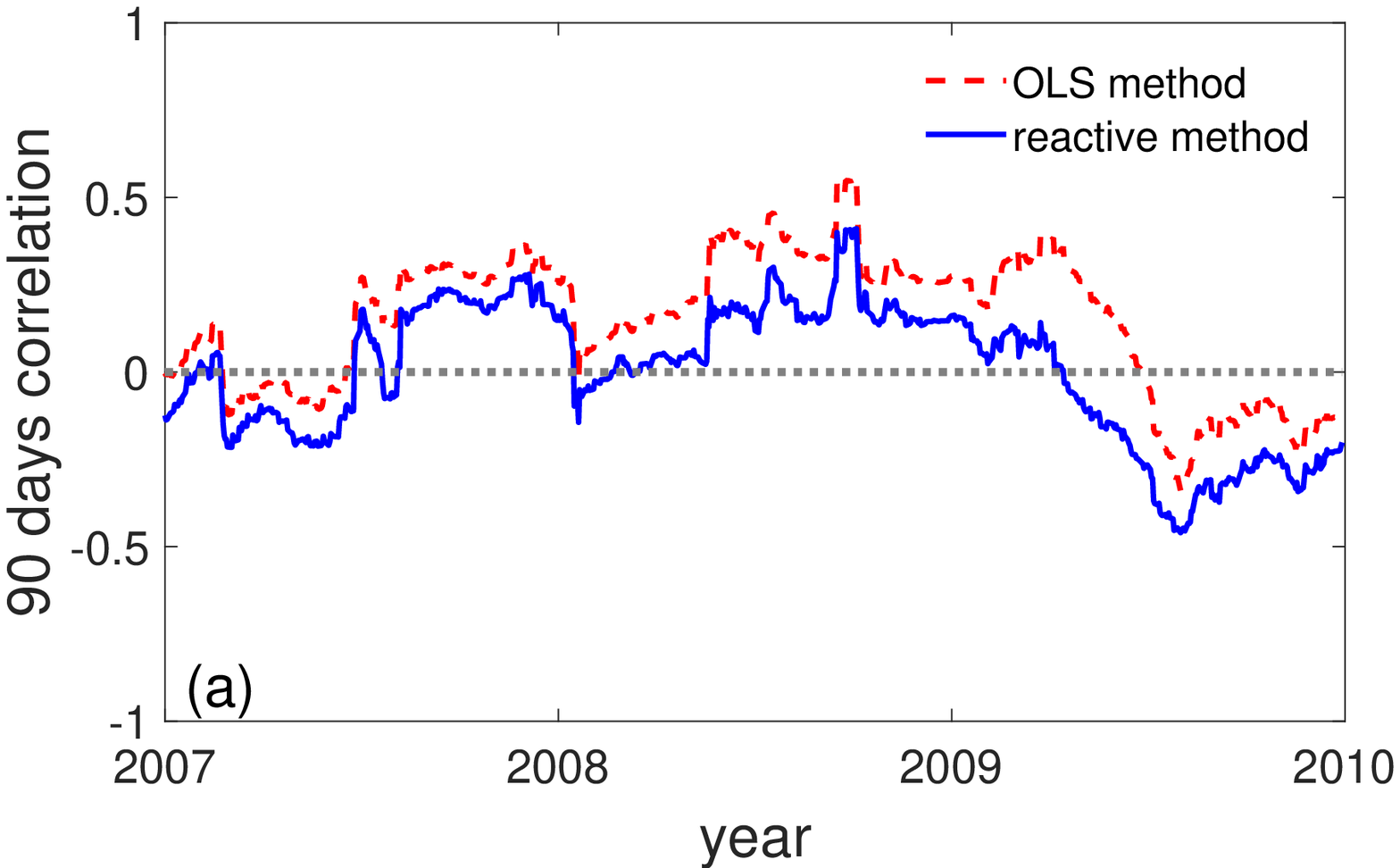}
\includegraphics[width=120mm]{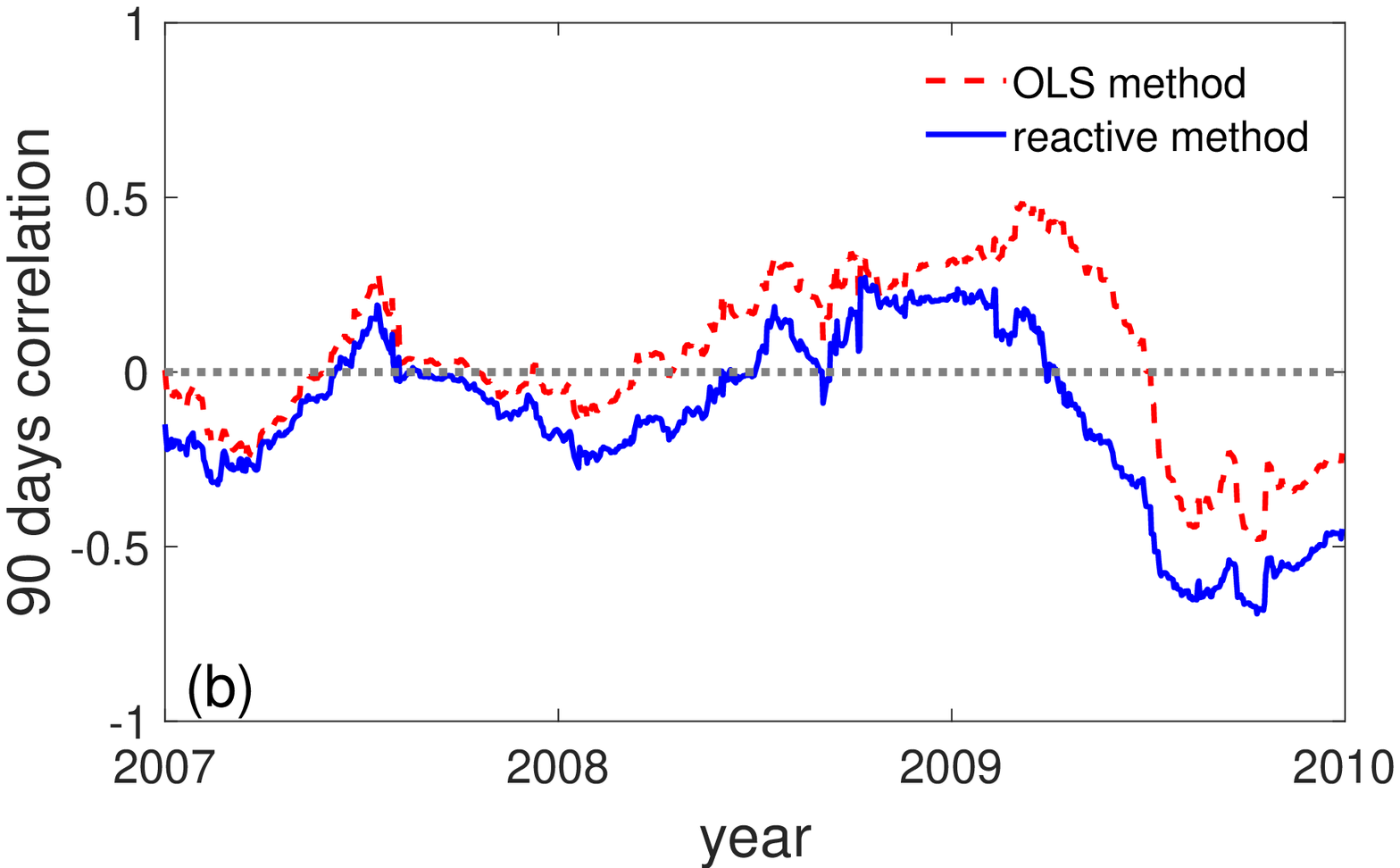}
\end{center}
\caption{
Ninety-day correlation of the short-term reversal factor with the
stock index in the European market {\bf (a)} and in the USA market
{\bf (b)}.  Solid and dashed lines present the proposed Reactive beta
model and the OLS methodology, respectively. A time frame surrounding
the financial crisis is chosen. Period: 2007-2010.}
\label{fig:momentum}
\end{figure}

We also illustrate these findings by presenting the correlation
between the stock index and the low volatility strategy (Figure
\ref{fig:betafactor}) and the short-term reversal strategy (Figure
\ref{fig:momentum}), which are the strategies with the highest bias.
A period surrounding the financial crisis was chosen (2007-2010).  One
can see that the beta, computed by the OLS, is highly positively
exposed to the stock index in 2008.  In turn, the exposure is reduced
within the reactive model.  The improvement becomes even more
impressive in extreme cases when the strategies are stressed by the
market.  We see that in some extreme cases (stress period with extreme
strategies), the common approach could generate high biases ($-50\%$
for the short-term reversal strategies in 2008-2009 and $-71\%$ for
the beta strategy in 2008).  In each case, our methodology allows one
to significantly reduce the bias.

\begin{table}
	\begin{center} 
		\begin{tabular}{|c|c|c|c|c|c|}   \hline
			&  Strategy $\backslash$ Method & \multicolumn{2}{c|}{OLS} & \multicolumn{2}{c|}{Reactive} \\    \hline
			\multirow{6}{3mm}[-1mm]{\begin{turn}{90}US\end{turn}} 
			
			&  Statistics : & Bias & CorSTD & Bias & CorSTD  \\    \hline
			&  low volatility	& -25.54\% & 21.73\% & -16.79\% &21.43\% \\
			&  short-term reversal	& 13.09\% &18.96\% &   -6.06\% &18.50\%  \\
			&   momentum  	&  -6.27\% &18.28\% &  -2.95\% &16.54\% \\
			&  size	&  -7.56\% &17.00\% &  -1.84\%&17.26\%  \\  \hline
			\multirow{6}{3mm}[4mm]{\begin{turn}{90}Europe\end{turn}} 
			&  low volatility	& -22.39\% &19.97\% & -14.68\% &20.94\% \\
			&  short-term reversal	&  13.05\% &17.51\% &  0.64\% &14.52\% \\
			&   momentum	&  -4.42\% &18.03\% &  -1.55\% &17.23\% \\
			&  size 	&   3.12\% &17.15\% &   3.79\% &15.63\%  \\  \hline
		\end{tabular}
	\end{center}
\caption{
Bias is defined as the correlation over the whole sample between
the stock index and each of the OLS and Reactive strategies for the US
and Europe markets. CorSTD is defined as the standard deviation of the
90-days correlation over the whole sample between the stock index and
each of the OLS and Reactive strategies for the US and Europe
markets.  The residual bias for the low volatility strategy in the
reactive method can be explained by the selection bias as demonstrated
in Appendix \ref{sec:selection}. Period: 2000-2015.}
\label{tab:schemes}
\end{table}

%%%  Beginning of red color 	

\section{Robustness Checks}
\label{sec:robustness}

This section presents robustness check analysis by comparing the
quality of several methods for beta measurements against the reactive
beta model.  We build the comparative analysis based on two important
articles in order to explore two aspects of the beta estimation.
\citet{Chan92} enable to assess robustness statistics of some
alternatives methods to classical ordinary least square (OLS) when
assuming implicitly that betas are static and returns are
homoscedastic.  This section extends their work by including
alternative dynamics beta estimators to be coherent with our reactive
model and with the work by \citet{Engle16} that demonstrates that the
betas are significantly time-varying using dynamic conditional betas.
The presentation of the models and methods are located in the Appendix
\ref{sec:methods}.

\subsection{Monte Carlo simulations}
	
In financial research, one often resorts to simulated data to estimate
the error of measurements.  For instance, \cite{Chan92} built their
main results on numerical simulation while applying real data for
simple comparison between betas estimated with OLS and quantile
regression (QR).

The comparative analysis is based on a two-step procedure.  The first
step simulates returns using different models that capture some
markets patterns and the second step estimates the beta from simulated
returns by using our reactive method and alternative methods.  We
tested the same estimators as used by \cite{Chan92} that includes the
OLS, the minimum absolute deviation (ABSD), and the Trimean quantile
regression (TRM).  We also added two variations of the dynamical
conditional correlation (DCC) which has become a mainstream model to
measure conditional beta when beta is stochastic
\citep{Bollerslev88,Bollerslev90,Engle02,Cappiello06}.  We
analyze the error of measurements that we defined as the difference
between the measured beta and the true beta of simulated data.

\subsubsection{The first step: simulation}
	
\begin{figure}
\begin{center}
\includegraphics[width=120mm]{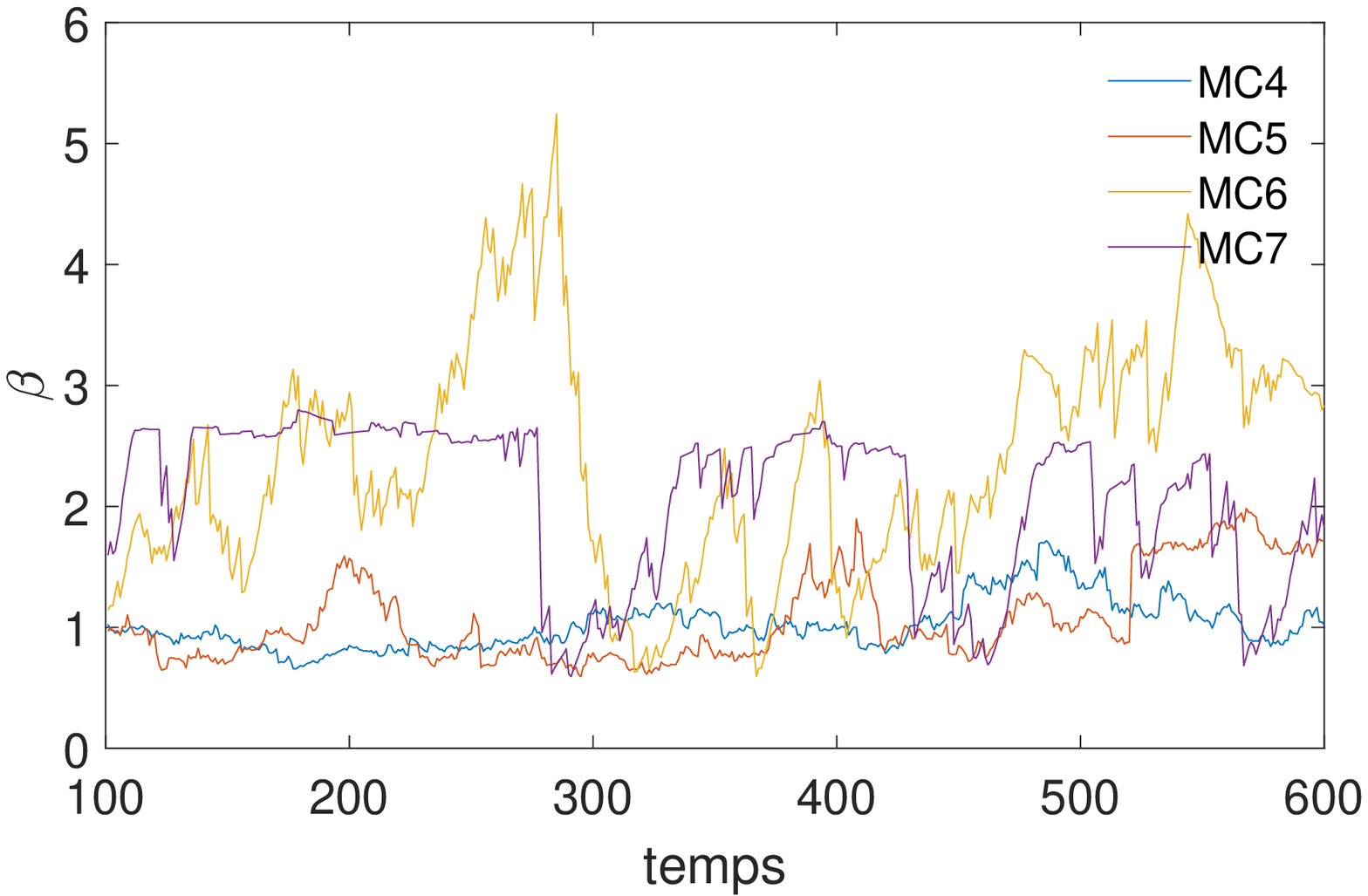} % {betasimSh.eps}
\includegraphics[width=60mm]{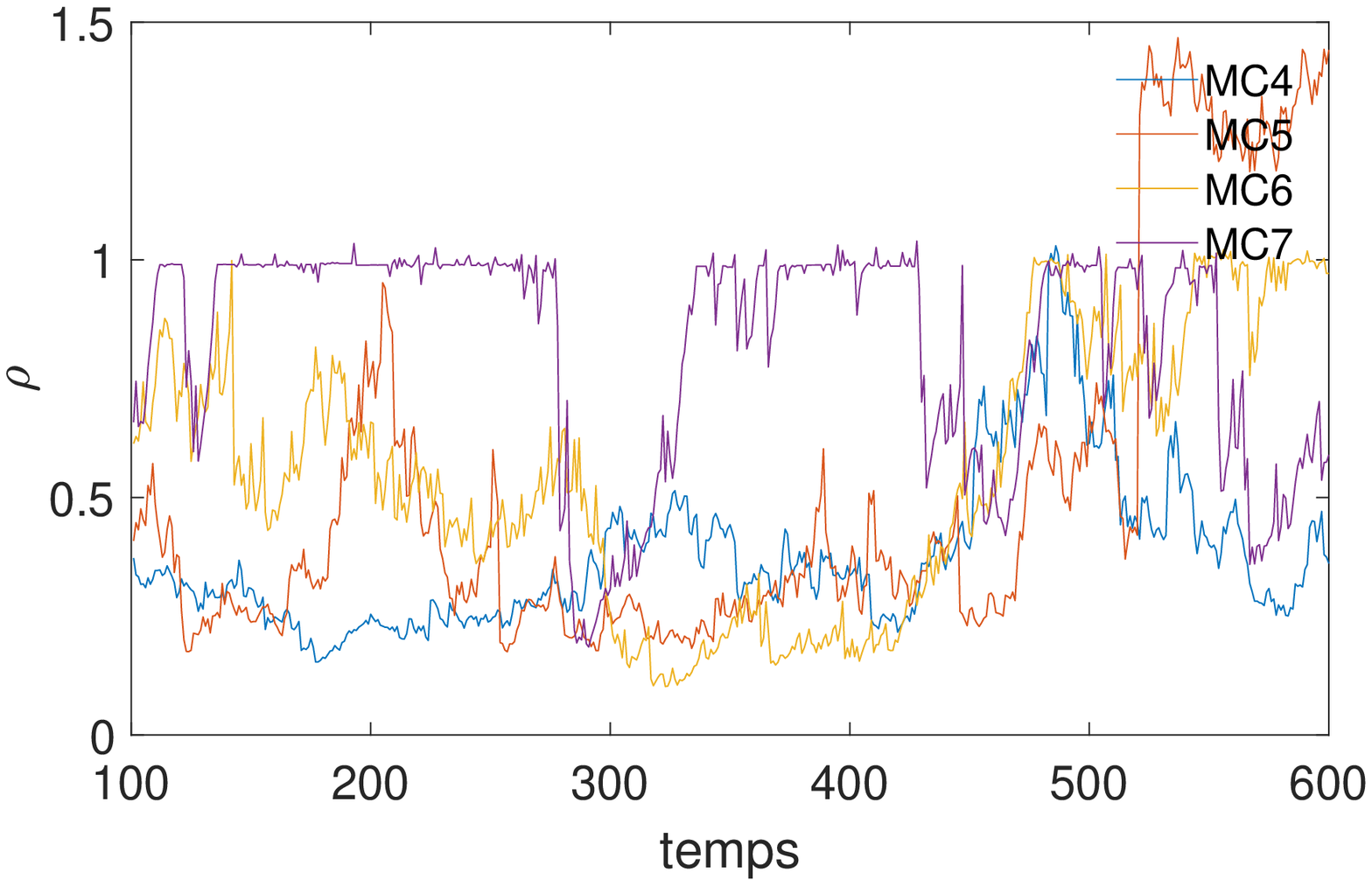} % {rhosimSh.eps}
\includegraphics[width=60mm]{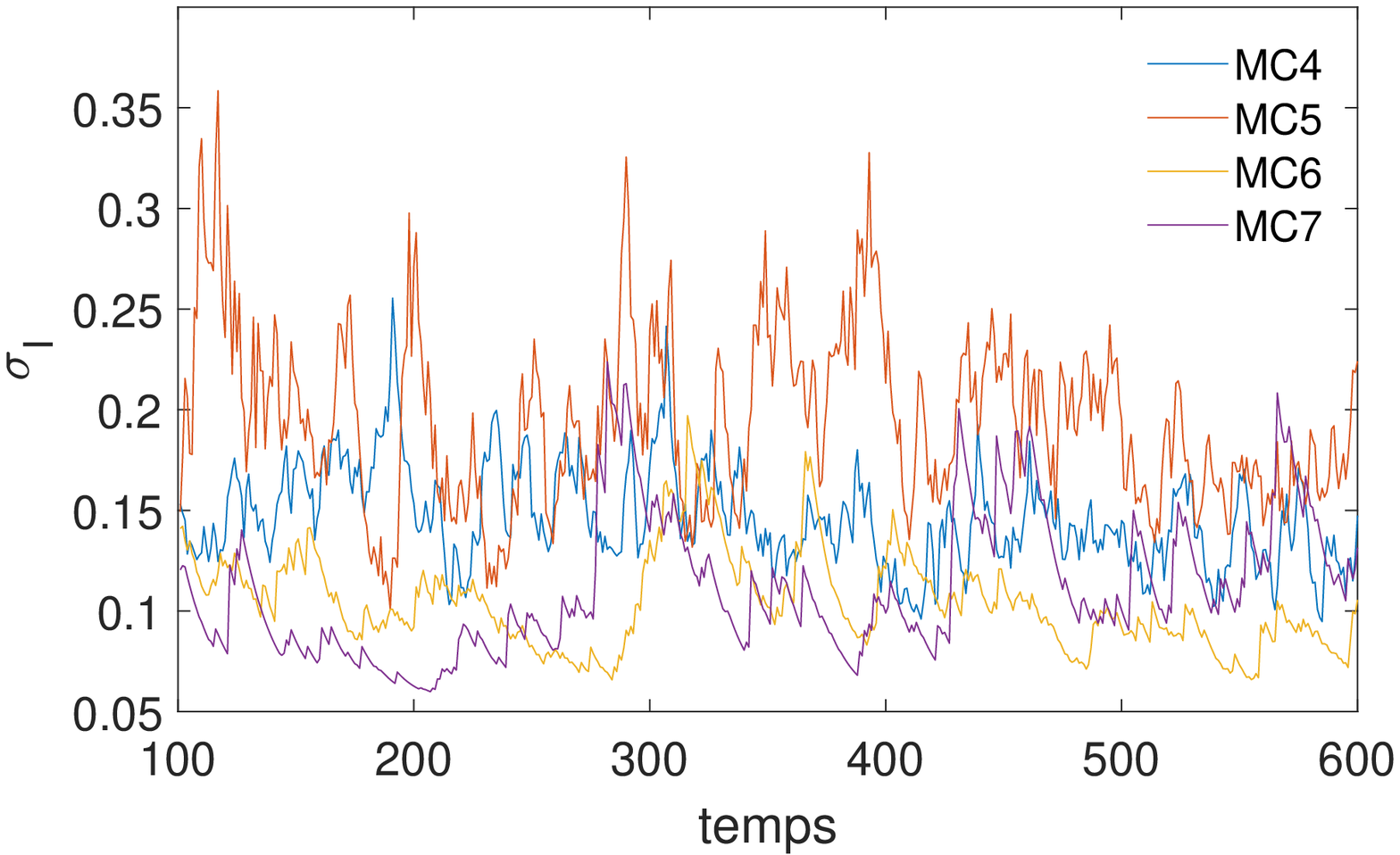} % {volindicesimSh.eps}
\includegraphics[width=60mm]{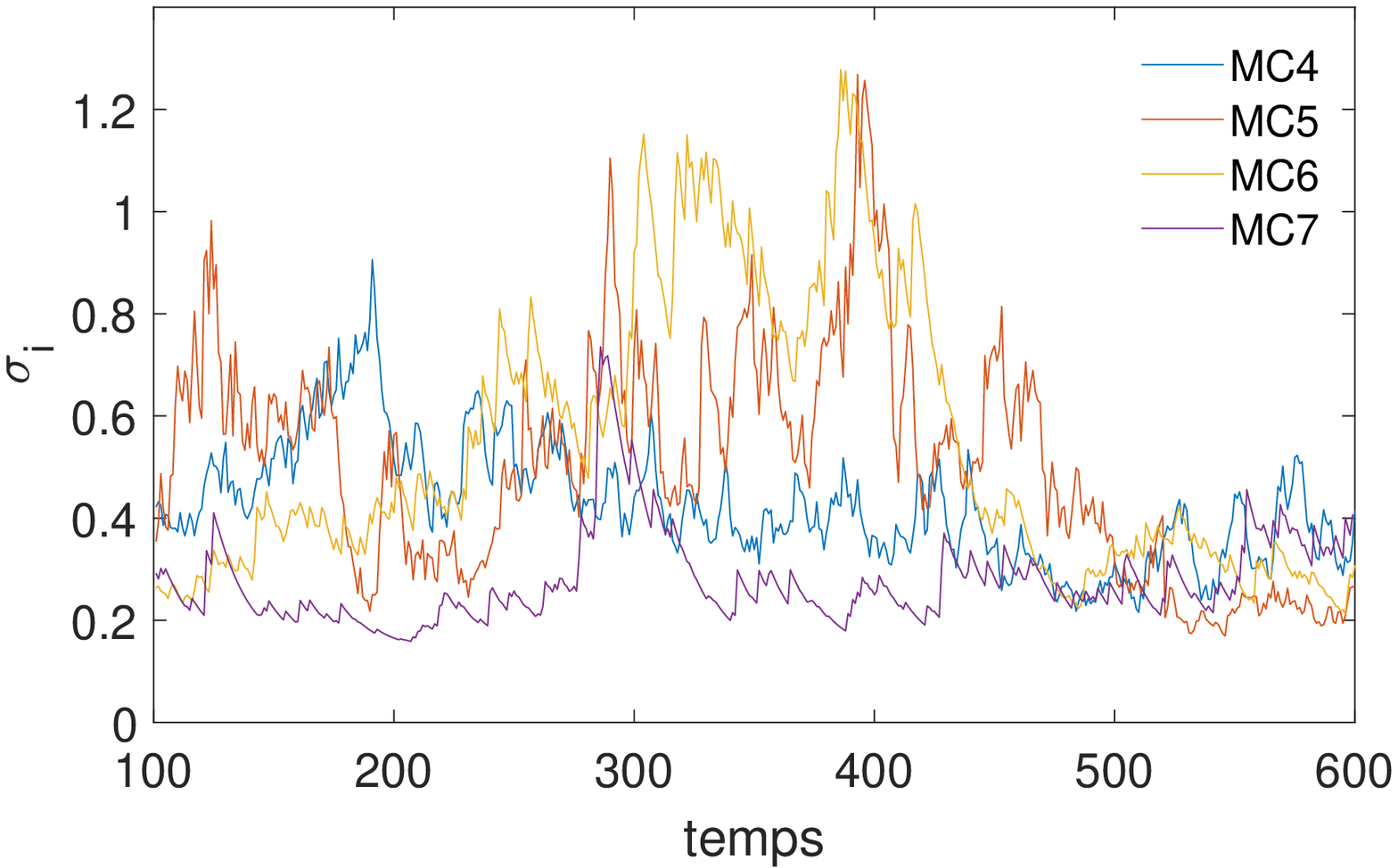} % {volstocksimSh.eps}
\includegraphics[width=60mm]{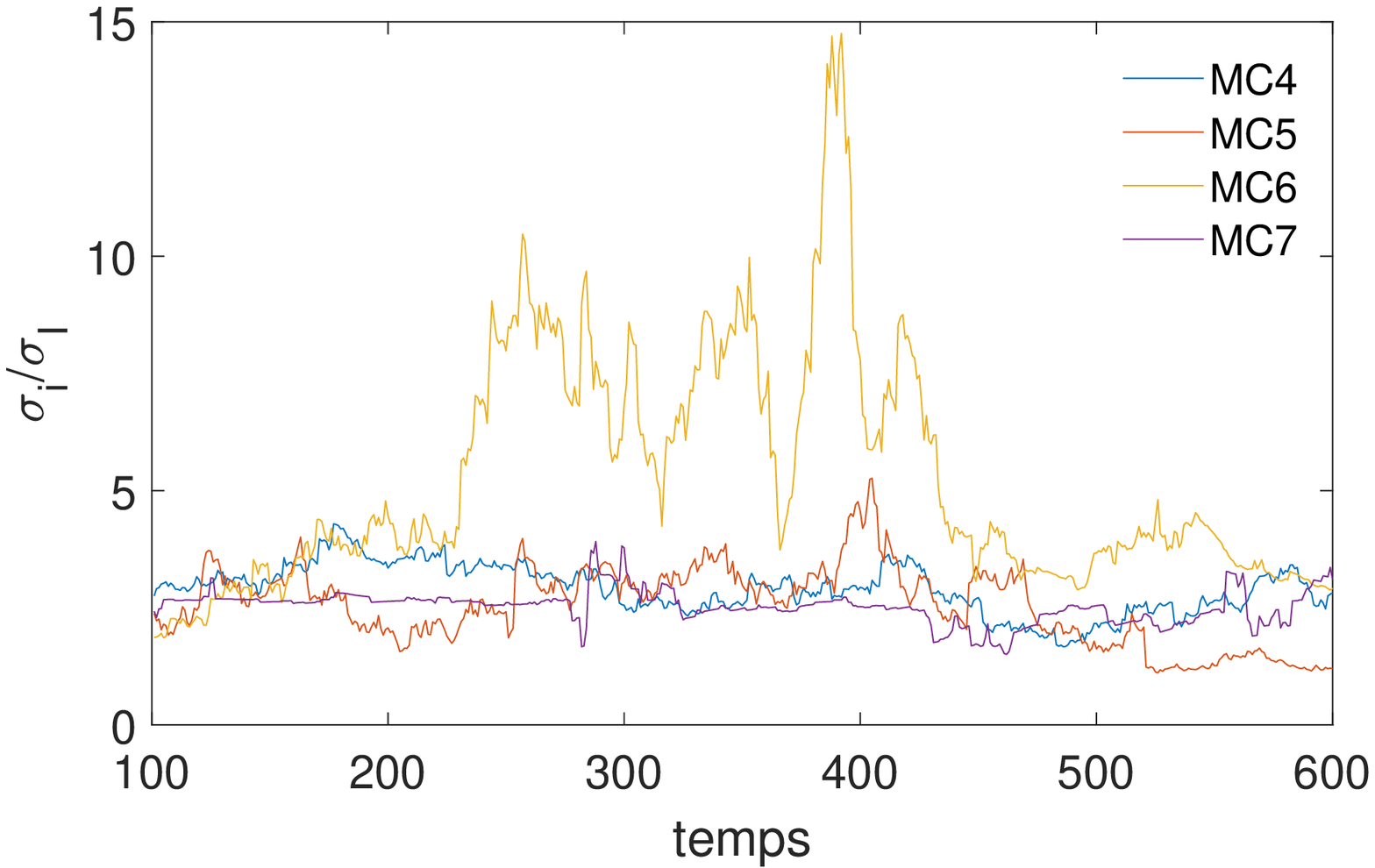} % {volrelativesimSh.eps}
\end{center}
\caption{
Simulated paths for the models MC4 -- MC7.  True conditional beta
(top), true conditional correlation (middle left), true conditional
stock index volatility (middle right), true conditional single stock
volatility (bottom left), true conditional relative volatility (bottom
right) are plotted.  Paths limited to 500 days, that are independent
from model to model, capture the same order of magnitude of variation
in volatilities, beta and correlation.  }
% JLI_Seb_RCM_fig1();
\label{fig:MC} 
\end{figure}

The first step simulates 30,000 paths of $T$=1,000 consecutive returns
for both the stock index and the single stock.  It allows also to
generate 1,000 conditional ``true'' expected beta per path
(Fig. \ref{fig:MC}).  To that end, following \cite{Chan92}, normally
distributed residuals and Student-t distributed residuals are
considered to take into consideration robustness of different methods
to outliers.

In our setting, we implemented seven Monte Carlo simulations for the
returns $r_i$ and $r_I$.  We targeted in simulations the realistic
case of an unconditional single stock annualized volatility of 40\%,
an unconditional stock index volatility of 15\% and an unconditional
beta of 1.  That is important to target the realistic correlation
between the index and the stock of 0.4.  Indeed the relative precision
of the beta measurement is inversely proportional to the square root
of the number of returns when correlation is close to zero.  First, we
consider the naive version of the market model, based on
Eq. (\ref{eq:simplemarketmodel}), that we call ``the basic market
model'' For the case of constant beta, as in paper by \cite{Chan92},
the simulated data are based on the hypothesis of a null intercept and
beta is equal to $1$ to characterize the ideal case with a Gaussian
(MC1) or a t-student distribution (MC2) for residuals.  In the most
simple reactive version of the market model that we call ``the
reactive market model'', normalized returns $\tilde r_i$ and $\tilde
r_I$ are first generated randomly through
Eq. (\ref{eq:simplereactivemodel}) with a normalized beta set to 1.
Then, based on the level $L_{s}$, $L_{is}$ that are respectively the
slow moving averages of the stock index and the stock prices defined
in Eq. (\ref{eq:Lslow}), we generate $\delta I$ and $\delta S$ defined
in Eq. (\ref{eq:tilder}), then $r_i$ and $r_I$, and finally update
$L_{s}$ and $L_{is}$ .  That model is sufficient to capture the
leverage effect on beta with increasing beta as soon as single stock
underperforms the stock index.  Even if the normalized beta is set to
unity (MC3 and MC4), the denormalized beta in
Eq. (\ref{eq:reactivevolatility}) becomes time dependent
(Fig. \ref{fig:MC}).  As previously, MC3 and MC4 differ by the
distribution of residuals, Gaussian (MC3) versus Student-t (MC4).

For the case of time-varying beta (MC 3 to 5), we used two versions of
the reactive market model in Eq. (\ref{eq:simplereactivemodel}): the
reduced version with only the leverage effect components that is
enough to generate stochastic beta in
Eq. (\ref{eq:reactivevolatility}), and the full version with all
components including the nonlinear beta elasticity.  For the full
version (MC5), we generated stochastic $\tilde{\sigma}_i$ and
$\tilde{\sigma}_I$ that generate $\tilde r_i$ and $\tilde r_I$ from
Eq. (\ref{eq:simplereactivemodel}) using the normalized beta fixed to
$\mathcal{F}(t)\mathcal{L}(t)$ (see definitions in Eqs. (\ref{eq:F})
and (\ref{eq:L})).  That allows to generate returns that capture the
leverage effect pattern and the empirical non-linear beta elasticity
(Fig. \ref{fig:beta_log} and Fig. \ref{fig:fdependingonbeta}).

For the case of time-varying beta (MC 6 to 7), we used another way to
generate random returns that capture a time-varying beta through the
implementation of the dynamic conditional correlation (DCC) model
\citep{Engle02}, which generalizes the GARCH(1,1) process to two
dimensions.  This is a mainstream model which has two variations:
symmetric and asymmetric, the latter capturing the leverage effect.
Symmetric and asymmetric versions of DCC model are denoted as MC6 and
MC7.

To summarize, seven Monte Carlo simulations:
\begin{itemize}
\item 
MC 1: The basic market model in Eq. (\ref{eq:simplereactivemodel})
where residuals ($\epsilon_i$) are normally distributed and constant
beta is set to 1.

\item 
MC 2: The basic market model in Eq. (\ref{eq:simplereactivemodel})
where residuals ($\epsilon_i$) follow a Student-t distribution (with
three degrees of freedom) and constant beta is set to 1.

\item 
MC 3: The reduced reactive market model in
Eq. (\ref{eq:simplereactivemodel}) where residuals ($\tilde
\epsilon_i$) are normally distributed with constant volatilities
($\tilde \sigma_i$, $\tilde \sigma_I$) and constant renormalized beta
($\tilde \beta$) set to 1 but the denormalized beta is now depending
on time (Fig \ref{fig:MC}).  The conditional beta ($\beta$) is now a
mean reversion process with a relaxation time $1/\lambda_s=50$ days.
MC3 uses only the leverage effect component but not the nonlinear beta
elasticity.

\item 
MC 4: The reduced reactive market model in
Eq. (\ref{eq:simplereactivemodel}) where residuals ($\tilde
\epsilon_i$) follow a Student-t distribution (with three degrees of
freedom) with constant relative volatility and constant renormalized
beta set to 1.
		
\item 
MC 5: The full reactive market model in
Eq. (\ref{eq:simplereactivemodel}) where residuals ($\tilde
\epsilon_i$) follow a Student-t distribution (with three degrees of
freedom) whose standard deviation ($s_i$) is stochastic and where the
normalized stock index return ($\tilde r_I$) is a Gaussian whose
standard deviation ($s_I$) is also stochastic.  We suppose that
$\log(s_I)$ and $\log(s_i)-\log(S_I)$ follow two independent
Ornstein-Uhlenbeck processes (with the relaxation time of 100 days and
the volatility of volatility of 0.04).  In that way the stock index
annualized volatility could jump up to 40\%.  Normalized beta, that
was set to 1 in MC4, is now set to $\mathcal{F}(t)\mathcal{L}(t)$ to
take into account the nonlinear beta elasticity (see definitions in
Eqs. (\ref{eq:F}) and (\ref{eq:L})).  Both leverage effect and
stochastic normalized volatilities make the beta defined in
Eq. (\ref{eq:beta_final}) )and volatilities time-depended (Fig.
\ref{fig:MC}).

\item 
MC 6: The symmetric DCC model in two dimensions, which generates
volatilities of volatilities and correlation of similar amplitude as
MC5 (Fig. \ref{fig:MC}).

\item 
MC 7: The ADCC model in two dimensions, which generates volatilities
of volatilities and correlation of similar amplitude as MC5
(Fig. \ref{fig:MC}).
		
\end{itemize}
	
In Fig. \ref{fig:MC}, we plot a Monte Carlo path generated for true
beta for MC 3 to 7 (MC1 and MC2 are excluded as they generate true
beta of 1).  We also plot the conditional correlation and volatilities
that are highly volatile and make the estimation of the conditional
beta complicated.

\subsubsection{The second step: measurements}
	
The second step is devoted to the analysis of the error measurement of
the beta estimations defined as the difference between the measured
beta and the true beta of simulated data.  In our setting, we test 5
alternative beta estimations that should replicate as close as
possible the true beta.  Notice that in all five configurations, we
use an exponentially weighted scheme to give more weight to recent
observations to be in line with the reactive market model
($1/\lambda_\beta=90$).  As a consequence, in a path of $T$=1,000
generated returns, only the 90 last returns really matters (note that
\cite{Chan92} based their statistics on 35 returns with an equal
weight scheme).
The first alternative method is the Ordinary Least Square (OLS) of the
returns which was also implemented in the empirical test based on real
data.  Note that the OLS would give the same measurement than our
reactive method if parameters were set differently ($\lambda_s=1$,
$\lambda_f=1$, $l=l'=0$, $f=0$).  The square errors in the OLS are
weighted by $(1-\lambda_\beta)^{T-t}$.
The second method estimates the beta by using the Minimum Absolute
Deviation (MAD) that is supposed to be less sensitive to outliers as
absolute errors instead of square errors are minimized.  The absolute
errors are weighted by $(1-\lambda_\beta)^{T-t}$.
The third alternative is the beta computed from the Trimean quantile
regression (TRM) that is reputed to be more robust to outliers
according to \cite{Chan92}.  The absolute errors are also weighted by
$(1-\lambda_\beta)^{T-t}$.
The fourth and fifth methods are the conditional beta computed from
the DCC model.  The DCC method was calibrated using the same
exponential $(1-\lambda_\beta)^{T-t}$ weights introduced in the
log-likelihood function to extract the optimal unconditional
volatilities and correlations, while other parameters such as the
relaxation time and volatilities of volatilities and volatilities of
correlations were set to the values that were used for Monte Carlo
simulation.

We summarize the reactive method and the five alternative methods that
were implemented to estimate the beta:
\begin{itemize}
\item $\beta_{OLS}$: beta estimated by the Ordinary Least Square method; 
		
\item $\beta_{MAD}$: beta estimated by the Minimum Absolute Deviation method;
		
\item $\beta_{TRM}$: beta estimated by the Trimean Quantile regression;
		
\item $\beta_{DCC}$: $T^{\rm th}$ conditional beta estimated by using the  DCC model;
			
\item $\beta_{ADCC}$: $T^{\rm th}$ conditional beta estimated by using the ADCC model;
		
\item $\beta_{R}$: beta estimated by the reactive method in Eq. (\ref{eq:beta_final}).
		
\end{itemize}

\subsubsection{The statistics}

We analyze for every path the error of measurement defined as the
difference between the measured beta based on different methods
applied to $T$ returns and the true value of beta at time $T$.

To assess the quality of different methods, we use three statistics
following \cite{Chan92}.
The first statistics is the bias and gives the average error of
measurement.  Yielding the bias is more informative than simply
yielding an estimated average estimation of beta as in our case the
theoretical expected simulated conditional beta is not always 1 but
fluctuates around 1 for time-varying models from MC3 to MC6.  As we
focused on capturing the leverage effect in the beta measurement we
also define winner (loser) stocks that are the stocks that have
outperformed (underperformed) the stock index during the last month.
Due to the leverage effect, the OLS method is expected to
underestimate beta for loser stocks and to overestimate beta for
winner stocks.  It would be interesting to see how robust is the
improvement of the reactive beta.  We therefore measure the average
error among the loser and winner stocks.  The loser and winner biases
are related to the bias in hedging of the short term reversal strategy
measured on real data and could confirm the robustness of the
empirical measurements. We also define the low (high) beta stocks that
are the stocks whose conditional true beta is lower (higher) than 1.
We measure the average error among low and high beta stocks that are
related to the bias in hedging of the low beta strategy measured from
real data and could confirm the robustness of the beta measurement
when adding the component describing the nonlinear beta elasticity.

The second statistics is the ABSolute Deviation (ABSD) of measurement.
It reflects the average absolute errors such that the positive and
negative sign errors cannot be mutually compensated.  It is a
measurement of the robustness.
The third statistics, that is equivalent to ABSD, is the inverse of
the variance of the errors of measurement ($\frac{V_{OLS}}{V_{m}}$) to
characterize the relative robustness of the alternative beta
estimation. The alternative beta method (with subscript $m$) that
brings the highest improvement is the one with the highest ratio.
	
The three statistics that were implemented are summarized:
\begin{itemize}
\item Statistics 1: the bias, the  winner bias and the loser bias, the low bias and the high bias;
\item Statistics 2: the absolute deviation of measurement (ABSD);
\item Statistics 3: the relative variance statistics $\frac{V_{OLS}}{V_{m}}$.
\end{itemize}

\subsection{Empirical tests}
	
We summarize statistics in Table \ref{tab:MC}.  We see that all
methods are unbiased on average in most Monte Carlo simulations.  But
this is misleading as biases from one group of stocks can be
significant and offset others.

\subsubsection{Winner and loser bias}

The estimated $\beta_{DCC}$ and $\beta_{ADD}$ appear to be biased as
soon as fat tails are included (MC2).

The reactive beta is the only one to be unbiased for winner and loser
stocks when the leverage effect is introduced in Monte Carlo (MC 3, 4,
5).  The biases for winner stocks and loser stocks are significant for
all methods except for the reactive beta.  The biases are amplified
when a fat tail of residuals distribution is introduced (MC 4).
Winner/loser biases can reach 14\%.  That is in line with the
empirical test implemented on real data where we see that the reactive
method reduces the bias of hedging of the short-term reversal strategy
(Tab. \ref{tab:schemes}).

When all components that deviate from the Gaussian market model are
mixed in MC5 (fat tails, nonlinear beta elasticity, stochastic
volatilities, leverage effect) we see a kind of cocktail effect as
bias is generated for most methods on average and not only in some
groups of stocks.  The reactive method provides the best results and
is the only method that has no bias.  $\beta_{MAD}$ and $\beta_{TRM}$
that were supposed to be robust appear to perform very badly with high
bias (average, loser or winner) as soon as stochastic volatility is
added that is confirmed with MC6 and MC7.

We also see that the reactive model looks to be incompatible with the
DCC or ADCC model.  Indeed MC5 generates high bias for $\beta_{DCC}$
and $\beta_{ADD}$ in the winner and loser stocks even if the leverage
effect and the dynamic beta are implemented in the ADCC.  In the same
way MC 6 generates bias for the reactive method that are even
amplified when leverage effect is generated through MC7.  We can
wonder which model is the most realist.  Both ADCC and the reactive
model capture the volatility clustering and leverage effect patterns
but their dynamics is in reality very different.  In the reactive
model, volatility increases as soon as price decreases, and decreases
as soon as price increases whereas ADCC needs to see its volatility
increase a negative return, higher than expected ($\gamma \left(
\sigma_i^{2} [\xi_i^-(t)]^2 -\tilde\sigma_i^{2}\right)>0$, see
Eqs. (\ref{eq:beta_aux1}, \ref{eq:beta_aux2})).  The reactive beta
model has its three components that were fitted to three well
identified effect (the specific leverage through the retarded effect,
the systematic leverage through the panic effect and the non linear
beta elasticity) whose main parameters appears to be stable and
universal for all markets.  \citet{Bouchaud01} measured most of the
parameters for 7 main stock indexes.  Relaxation time is around 1 week
for the panic effect ($\lambda_s=0.1484$), relaxation is 40 days for
the retarded effect ($\lambda_s = 0.0241$), the leverage parameter for
the panic effect is $l= 8$.  The systematic leverage parameter
$\ell-\ell^{\prime} = 0.91$ was the only one to have been measured
through the implied correlation only from the US market.  The
parameters of the beta elasticity were measured similar for both the
European and the US market.  The different thresholds are $0.5$ and
$1.6$ in beta of the non linear beta elasticity separating low beta
stocks from speculative stocks).  Parameters $a$, $b$, $\gamma$,
$a_{\rho}$, $b_{\rho}$, $\gamma_{\rho}$ of the DCC and ADCC were based
on the work by \cite{Sheppard17} but $b$ and $b_{\rho}$ which are the
``decay coefficients'', describe relaxation times of 10 days and 13
days that are different from those used in the reactive volatility
model.

\subsubsection{High and low beta bias}

The reactive beta is the only one that reduces the bias for low and
high beta stock when stochastic volatility is introduced and when the
empirical nonlinear beta elasticity is implemented (MC 5).  That is in
line with the empirical test implemented on real data where we see
that the reactive method reduces the bias of hedging of the low
volatility strategy (Tab. \ref{tab:schemes}).

\subsubsection{ABSD and $V_{OLS}/V_{m}$}

The $\beta_{OLS}$, that is the theoretical optimal estimation for
Monte Carlo simulated returns with the Gaussian market model (MC1),
gives similar statistics to that of the reactive beta for the MC3.  In
this case (MC3), the reactive method outperforms the other considered
methods.
%
% which is the optimal
%estimation with the lowest ABSD and the highest $V_{OLS}/V_{m}$
%statistics for the Monte Carlo with leverage effect but without fat
%tails (MC3).  Statistics are similar to those of ).  
The ABSD of 0.17 is entirely explained by irreducible statistical
noise that is intrinsic to any regression based on approximately 90
points with a weak correlation.

When a fat tail is incorporated to the residual (MC4), the ABSD of the
reactive beta is increased and becomes intermediate between the ABSD
of $\beta_{OLS}$, $\beta_{MAD}$ and $\beta_{TRM}$.  $\beta_{MAD}$ and
$\beta_{TRM}$ are more robust in presence of fat tails.  The reactive
beta is expected to be as sensitive as the OLS would be due to the
outliers.  The reactive method could be still improved if a TRM
regression was implemented instead of the classical OLS to measure the
normalized beta between normalized returns.  When stochastic
volatility and correlation are introduced (MC5, MC6 and MC7), the
reactive beta becomes as robust as $\beta_{MAD}$ and $\beta_{TRM}$
based on ABSD.

\begin{table}[h]
%\footnotesize
\scriptsize
\centering
%	\begin{center}  
%\resizebox{\textwidth}{!}{%
		\begin{tabular}{ |c|c|c|c|c|c|c|c| }	\hline
			Method &  Bias & Winner Bias& Loser Bias& Low Bias& High Bias & ABSD   & Vols/Vm \\ \hline\hline
			\multicolumn{6}{c}{	MC1 Gaussian basic market model} \\ \hline			
			$\beta_{OLS}$&-0.00&-0.00&-0.00&&&0.16&1.00\\ \hline
			$\beta$ Reactive&0.00&-0.05*&0.05*&&&0.18&0.79\\ \hline
			$\beta_{DCC}$&0.04*&0.05*&0.03*&&&0.23&0.51\\ \hline
			$\beta_{ADCC}$&0.09*&0.01&0.17*&&&0.25&0.44\\ \hline
			$\beta_{MAD}$&-0.00&0.00&-0.01&&&0.20&0.65\\ \hline
			$\beta_{TRM}$&-0.00&0.00&-0.01&&&0.20&0.68\\ \hline			
			\multicolumn{6}{c}{	MC2  t-Student basic market model} \\ \hline			
			$\beta_{OLS}$&-0.00&0.01&-0.01&&&0.28&1.00\\ \hline
			$\beta$ Reactive&0.01&-0.06*&0.08*&&&0.31&0.82\\ \hline
			$\beta_{DCC}$&0.13*&0.14*&0.12*&&&0.39&0.67\\ \hline
			$\beta_{ADCC}$&0.25*&0.15*&0.35*&&&0.46&0.57\\ \hline
			$\beta_{MAD}$&-0.00&-0.00&-0.00&&&0.22&2.18\\ \hline
			$\beta_{TRM}$&-0.00&-0.00&-0.00&&&0.22&2.24\\ \hline			
			\multicolumn{6}{c}{	MC3  Gaussian reduced reactive market model} \\ \hline
			$\beta_{OLS}$&-0.00&0.07*&-0.07*&0.07*&-0.07*&0.19&1.00\\ \hline
			$\beta$ Reactive&-0.00&0.02*&-0.02*&0.02*&-0.02*&0.17&1.27\\ \hline
			$\beta_{DCC}$&0.04*&0.10*&-0.02&0.11*&-0.02&0.24&0.62\\ \hline
			$\beta_{ADCC}$&0.09*&0.06*&0.12*&0.07*&0.11*&0.24&0.66\\ \hline
			$\beta_{MAD}$&-0.01&0.06*&-0.08*&0.06*&-0.08*&0.22&0.73\\ \hline
			$\beta_{TRM}$&-0.01&0.06*&-0.08*&0.06*&-0.08*&0.22&0.75\\ \hline
			\multicolumn{6}{c}{	MC4  t-Student reduced reactive market model} \\ \hline
			$\beta_{OLS}$&0.01&0.13*&-0.11*&0.12*&-0.10*&0.35&1.00\\ \hline
			$\beta$ Reactive&-0.01&0.02&-0.04*&0.03&-0.05*&0.31&1.30\\ \hline
			$\beta_{DCC}$&0.12*&0.22*&0.02&0.27*&-0.01&0.47&0.84\\ \hline
			$\beta_{ADCC}$&0.26*&0.24*&0.28*&0.30*&0.21*&0.52&0.83\\ \hline
			$\beta_{MAD}$&-0.03*&0.09*&-0.14*&0.10*&-0.14*&0.27&2.68\\ \hline
			$\beta_{TRM}$&-0.03*&0.09*&-0.14*&0.10*&-0.14*&0.27&2.76\\ \hline
			\multicolumn{6}{c}{	MC5  t-Student full reactive market model} \\ \hline
			$\beta_{OLS}$&-0.01&0.13*&-0.14*&0.14*&-0.22*&0.50&1.00\\ \hline
			$\beta$ Reactive&-0.04*&-0.00&-0.07*&0.05*&-0.17*&0.41&1.42\\ \hline
			$\beta_{DCC}$&-0.01&0.10*&-0.12*&0.20*&-0.32*&0.52&1.31\\ \hline
			$\beta_{ADCC}$&0.10*&0.10*&0.11*&0.29*&-0.17*&0.54&1.32\\ \hline
			$\beta_{MAD}$&-0.09*&0.04*&-0.22*&0.09*&-0.37*&0.38&2.43\\ \hline
			$\beta_{TRM}$&-0.09*&0.04*&-0.22*&0.09*&-0.36*&0.37&2.46\\ \hline
			\multicolumn{6}{c}{	MC6  Gaussian symmetric DCC model} \\ \hline
			$\beta_{OLS}$&-0.11*&-0.10*&-0.11*&0.06*&-0.27*&0.32&1.00\\ \hline
			$\beta$ Reactive&-0.07*&-0.11*&-0.02&0.09*&-0.23*&0.33&0.93\\ \hline
			$\beta_{DCC}$&-0.01&-0.00&-0.02*&-0.01&-0.01&0.16&4.09\\ \hline
			$\beta_{ADCC}$&0.02*&-0.08*&0.12*&0.05*&-0.01&0.22&2.06\\ \hline
			$\beta_{MAD}$&-0.14*&-0.13*&-0.15*&0.04*&-0.32*&0.34&0.89\\ \hline
			$\beta_{TRM}$&-0.14*&-0.13*&-0.15*&0.04*&-0.32*&0.34&0.90\\ \hline
			\multicolumn{6}{c}{	MC7  Gaussian asymmetric DCC model} \\ \hline			
			$\beta_{OLS}$&-0.09*&0.03&-0.24*&0.09*&-0.25*&0.30&1.00\\ \hline
			$\beta$ Reactive&-0.07*&0.02&-0.17*&0.10*&-0.21*&0.27&1.21\\ \hline
			$\beta_{DCC}$&-0.04*&0.04*&-0.15*&-0.00&-0.08*&0.21&2.08\\ \hline
			$\beta_{ADCC}$&-0.01&-0.01&-0.01&-0.00&-0.01&0.15&3.74\\ \hline
			$\beta_{MAD}$&-0.13*&-0.02&-0.28*&0.06*&-0.29*&0.32&0.92\\ \hline
			$\beta_{TRM}$&-0.13*&-0.01&-0.28*&0.06*&-0.29*&0.32&0.92\\ \hline			
			\hline
		\end{tabular}%}
%	\end{center}
\caption{
Monte-Carlo robustness test.  Statistics are provided for seven Monte
Carlo simulations and six different methods to estimate the beta.  We
estimated the statistics such as the bias that is the average error of
beta measurement; winner/loser biases are the biases among
winner/loser stocks.  Low/High biases are the biases among low/high
beta stocks. * indicates a bias superior to 3 standard deviation.
ABSD is the average of the error in absolute value.  Vols/Vm is the
variance of the error in the OLS case divided by the variance of the
error.  } 
\label{tab:MC}
\end{table}

%%%%%%%
	
%\pagebreak
%%%%%%%%%%%%%%%%%%%%%%%%%%%%%%%%%%%%%%%%%%%%%%%%%%
	
\section{Open problems in other fields}
\label{sec:application}
	
The estimated beta is used in a wide range of financial applications,
which includes security valuation, asset pricing, portfolio management
and risk management.  This extends also to corporate finance in many
applications such like financing decisions to quantify risk associated
with debt, equity and asset and for firm valuation when discounting
cash-flows using the weighted average cost of capital.
The most likely reason is that the beta describes systematic risk that
could not be diversified and that should should be remunerated.
However as explained, the OLS estimator of beta is subject to
measurement errors, which include the presence of outliers, time
dependence, the leverage effect, and the departure from normality.
%
%Indeed, the standard CAPM approach relies on implausible assumptions
%regarding the investor's preferences which contaminates the subsequent
%beta measurement.

\subsection{Asset Pricing}
	
\cite{Bali17} apply the DCC model by \cite{Engle16} to assess the cross-sectional
variation in expected stock returns.
% instead of the traditional
%unconditional beta derived from the CAPM.  
They estimate the conditional beta for the S\&P 500 using daily data
for each year from 1963 to 2009.  They test if the betas have
predictive power for the cross-section of individual stocks returns
over the next on to five days.  They show that there is no link
between the unconditional beta and the cross-section of expected
returns.  Most remarkably, they also show that the time-varying
conditional beta is priced in the cross-section of daily returns.  At
the portfolio level, they indicate that a long-short trading strategy
buying the highest conditional beta stocks and selling the lowest
conditional beta stocks yields average returns of 8\% per year.  So
conditional CAPM is empirically valid whereas unconditional CAPM is
empirically not valid.
% \citep{Bali17}.  
Moreover they showed that conditional beta when comparing to
unconditional beta would not have significant pricing effect on major
anomalies (size, book, momentum,...).  So one can see that DCC greatly
improves the empirical validation of the CAPM but does not change
pricing of anomalies.  We expect that the reactive method can bring
further improvements.  According to our robustness tests in
Sec. \ref{sec:robustness}, the leverage effect and the nonlinear beta
elasticity could also generate bias in the DCC estimation.  As our
reactive method was designed to correct for these biases, its use can
help to reveal pricing effects of the dynamic beta on major anomalies.
This point is an interesting perspective for future research.

\subsection{Corporate Finance}

To determine a fair discount rate for valuing cash-flows, the firm's
manager must select the appropriate beta of the project given that the
discount rate remains constant over time while the project may exhibit
significant variation over time and leverage effect due to the
debt-to-equity ratio.  As such, \cite{Ang04} discuss how to discount
cash-flows with time-varying expected returns in traditional set-up.
For instance, the traditional dividend discount model assumes that the
expected return along with the expected rate of cash-flow growth are
set constant while they are time-varying and correlated.  In practice,
in the first step, the manager computes the expected future cash-flows
from financial forecasts and then in a second step, the manager uses a
constant discount rate, usually relying on the CAPM to discounting
factor.  In contrast, \cite{Ang04} derive a valuation formula that
incorporates correlation between stochastic cash-flows, betas and risk
premiums.  They show that the greater the magnitude of the difference
between the true discount rates and the constant discount rate, the
greater the project's misvaluation.  They even show that when
computing perpetuity values from the discounting model, the potential
mispricing can even get worse.  They conclude that accounting for
time-varying expected returns can lead to different prices from using
a constant discount rate from the traditional unconditional CAPM.  The
impact of the leverage effect and of the non-linear elasticity of beta
on potential mispricing should be investigated.

%%%  End of red color	

\section{Conclusion}
\label{sec:conclusion}

We propose a reactive beta model with three components that account
for the specific leverage effect (when a stock underperforms, its beta
increases), the systematic leverage effect (when a stock index
declines, correlations increase), and beta elasticity (when relative
volatility increases, the beta increases).  The three components were
fitted and incorporated through elaborate statistical measurements.
An empirical test is run from 2000 to 2015 with exhaustive data sets
including both American and European securities.  We compute the bias
in hedging the most popular market neutral strategies (low volatility,
momentum and capitalization) using the standard approach of the beta
measurement and the reactive beta model.  Our main findings emphasize
the ability of the reactive beta model to significantly reduce the
biases of these strategies, particularly during stress periods.
Robustness check confirms that the reactive beta is not biased when
the leverage effect and beta elasticity are introduced and appear to
be robust when volatility of volatility and volatility of correlation
are introduced.  
%The reactive model has a very different behavior than
%the mainstream dynamic beta model.  

\newpage

%%%%%%%%%%%%%%%%%%%%%%%%%%%%%%%%%%%%%%%%%%%%%%%%%%%%%%%%%%%%%%%%%%%%
\appendix
\section{Selection bias}
\label{sec:selection}

Here, we provide some evidence that the bias in beta of the low
volatility factor comes from the selection bias: selection of the
bottom beta stocks yields the stocks whose beta is underestimated.

The measured beta $\beta_{im}$ of stock $i$ is obtained by a
standard linear regression of the $i$-th stock returns, $r_i$, to the
stock index returns, $r_I$,
\begin{equation}  \label{eq:ri_epsiloni}
r_{i}=\beta_{im} r_I+ \epsilon_i ,
\end{equation}
where $\epsilon_i$ is the residual return.  We suppose that the
measured beta of the stock $i$, $\beta_{im}$, is affected by noise,
\begin{equation}  \label{eq:beta_im}
\beta_{im}=\beta_{iT}+\eta_i ,
\end{equation}
where $\beta_{iT}$ is the true beta (which is unknown), and $\eta_i$
is the error of the measurement inherent to the linear regression.
The standard deviation of $\eta_i$, $\sigma_{\eta_i}$, depends on the
average correlation between the single stock $i$ and the stock index
and on the number $n$ of independent points used for the regression (which we
set at $n = \frac{1}{\lambda_{\beta}}=90$):
\begin{equation}
\sigma_{\eta_i}=\frac{\sigma_{\epsilon i}}{\sigma_I} \, \frac{1}{\sqrt{n}} ,
\end{equation}
where $\sigma_{\epsilon i}$ is the standard deviation of the residual
returns $\epsilon_{i}$.  Averaging the above relation over all stocks,
we obtain
\begin{equation}
\sigma_{\eta} = \frac{\langle\sigma_{\epsilon i} \rangle}{\sigma_I}\sqrt{\lambda_\beta} ,
\end{equation}
where $\langle \sigma_{\epsilon i}\rangle$ denotes the average.
According to Eq. (\ref{eq:ri_epsiloni}), the standard deviation of the
stock returns, $\sigma_i$, is
\begin{equation}
\sigma_i = \sqrt{\beta_{im}^2 \sigma_I^2 + \sigma_{\epsilon i}^2} \approx \sigma_{\epsilon i} ,
\end{equation}
because $(\beta_{im} \sigma_I /\sigma_i)^2 \ll 1$ (stocks are much
more volatile than the index).  We thus obtain
\begin{equation}  \label{eq:sigma_eta}
\sigma_{\eta} \approx \frac{\langle\sigma_i \rangle}{\sigma_I}\sqrt{\lambda_\beta}.
\end{equation}

The low volatility factor is 50\% long of the 30\% top $\beta_{im}$
stocks and 50\% short of the 30\% bottom $\beta_{im}$ stocks (here, we
consider only one sector for simplicity).  We adjust the most volatile
leg to target a beta neutral factor if we suppose that $\eta_i$ are
null.  In reality, when taking into account the difference between the
measured and the true beta, we obtain the beta of the low volatility
factor as:
\begin{equation}  \label{eq:beta_low}
\beta_{\textrm{low factor}}= -50\%\langle\beta_{iT}\vert i\in \textrm{Bottom}\rangle + 50\% \frac{\langle\beta_{im}\vert i\in \textrm{Bottom}\rangle }
{\langle\beta_{im}\vert i\in \textrm{Top}\rangle}\langle\beta_{iT}\vert i\in \textrm{Top}\rangle .
\end{equation}
This is essentially the beta neutral condition that we impose when
constructing the factor (see Appendix \ref{sec:Afactors}).  Here,
$\langle\beta_{im}\vert i\in \textrm{Bottom}\rangle$ is the average of the
measured beta over the stocks $i$ in the 30\% bottom in the {\it
measured} beta values $\beta_{im}$ (similar for other averages).

Defining $\Delta \beta_B$ and $\Delta \beta_T$ as
\begin{equation}
\langle\beta_{iT}\vert i\in \textrm{Bottom}\rangle=\langle\beta_{im}\vert i\in \textrm{Bottom}\rangle+\Delta \beta_B ,
\end{equation}
\begin{equation}
\langle\beta_{iT}\vert i\in \textrm{Top}\rangle=\langle\beta_{im}\vert i\in \textrm{Top}\rangle+\Delta \beta_T ,
\end{equation}
we rewrite Eq. (\ref{eq:beta_low}) as
\begin{eqnarray} \nonumber
\beta_{\textrm{low factor}} &=& -50\%\left(\langle\beta_{im}\vert i\in \textrm{Bottom}\rangle+\Delta\beta_B\right)
+ 50\% \frac{\langle\beta_{im}\vert i\in \textrm{Bottom}\rangle }{\langle\beta_{im}\vert i\in \textrm{Top}\rangle}
\left(\langle\beta_{im}\vert i\in \textrm{Top}\rangle+\Delta\beta_T\right) \\
&=& -50\% \Delta\beta_B + 50\% \frac{\langle\beta_{im}\vert i\in \textrm{Bottom}\rangle }{\langle\beta_{im}\vert i\in \textrm{Top}\rangle}
\Delta\beta_T .
\end{eqnarray}
Given that $\langle\beta_{im}\vert i\in \textrm{Bottom}\rangle \ll
\langle\beta_{im}\vert i\in \textrm{Top}\rangle$ (as the $\beta_{im}$
in the top quantile are higher than the $\beta_{im}$ in the bottom
quantile), we obtain the following approximation
\begin{equation}
\beta_{\textrm{low factor}}\approx-50\%\Delta\beta_B .
\end{equation}
If one knew the true $\beta_{iT}$ values and used them for
constructing the low volatility factor, the excess $\Delta \beta_B$
would be zero.  However, the true values are unknown, and one uses the
measured beta $\beta_{im}$ that creates a selection bias and the
nonzero $\Delta \beta_B$, as shown below.

To estimate $\Delta\beta_B$, we consider the true beta $\beta_{iT}$
and the measurement error $\eta_i$ as independent random variables and
replace the average over stocks by the following conditional
expectation
\begin{equation}  \label{eq:Delta_beta}
\Delta \beta_B = \langle\beta_{iT} - \beta_{im} \vert i\in \textrm{Bottom}\rangle 
\approx\E \{ \beta_{iT} - \beta_{im} | i \in \textrm{Bottom} \} = B.
\end{equation}
We have, then,
\begin{eqnarray} \nonumber
- B &=& \E \{ \eta_i | i \in \textrm{Bottom} \} = 
\int\limits_{-\infty}^\infty \eta \, \P \{ \eta_i \in (\eta,\eta+d\eta) | i \in \textrm{Bottom} \} \\
&=& \int\limits_{-\infty}^\infty \eta \, \frac{\P \{ \eta_i \in (\eta,\eta+d\eta), ~ i \in \textrm{Bottom} \}}
{\P \{ i \in \textrm{Bottom} \}} , 
\end{eqnarray}
where we wrote explicitly the conditional probability.  The
denominator is precisely the threshold determining the bottom
quantile, $\P \{ i \in \textrm{Bottom} \} = p$, which we set to
$30\%$.  We thus obtain
\begin{equation} 
- B = \frac{1}{p} \int\limits_{-\infty}^\infty \eta \, \P \{ \eta_i \in (\eta,\eta+d\eta) , ~ \beta_{im} - \beta_0 < Q \}, 
\end{equation}
where the event $i \in \textrm{Bottom}$ is equivalently written as
$\beta_{im} < \beta_0 + Q$, where $Q$ is the value of the measured
beta that corresponds to the quantile $p$, and $\beta_0$ is the mean
of $\beta_{im}$.  Using Eq. (\ref{eq:beta_im}) and the assumption that
$\beta_{iT}$ and $\eta_i$ are independent, one obtains
\begin{eqnarray}  \nonumber
- B &=&  \frac{1}{p} \int\limits_{-\infty}^\infty \eta \, \P \{ \eta_i \in (\eta,\eta+d\eta) , ~ \beta_{iT} - \beta_0 < Q - \eta \} \\
&=& \frac{1}{p} \int\limits_{-\infty}^\infty \eta \, \P \{ \eta_i \in (\eta,\eta+d\eta)\}\, \P\{ \beta_{iT} - \beta_0 < Q - \eta \} .
\end{eqnarray}

To obtain some quantitative estimates, we make a strong
assumption that both $\beta_{iT}$ and $\eta_i$ are Gaussian variables,
with means $\beta_0$ and $0$ and standard deviations $\sigma_\beta$
and $\sigma_\eta$, respectively.  We then obtain
\begin{equation} 
- B = \frac{1}{p} \int\limits_{-\infty}^\infty d\eta \, \eta \, \frac{\exp(-\eta^2/(2\sigma_\eta^2))}{\sqrt{2\pi} \, \sigma_\eta} 
\Phi\bigl((Q - \eta)/\sigma_\beta\bigr) ,
\end{equation}
where
\begin{equation}
\Phi(x) = \int\limits_{-\infty}^x dy\, \frac{e^{-y^2/2}}{\sqrt{2\pi}}
\end{equation}
is the cumulative Gaussian distribution.  Changing the integration
variable, one obtains
\begin{equation} 
- B = \frac{\sqrt{2} \sigma_\eta}{p \sqrt{\pi}} \int\limits_{-\infty}^\infty dx \, x \, \exp(-x^2) 
\Phi\bigl((Q - x \sqrt{2}\sigma_\eta)/\sigma_\beta\bigr) .
\end{equation}
Integrating by parts and omitting technical computations, we obtain
\begin{equation} 
B = \frac{\sqrt{2} \sigma_\eta}{p \sqrt{\pi}} \, \frac{\sigma_\eta}{2\sigma_\beta \sqrt{1+b^2}} 
\exp\biggl(-\frac{a^2}{1+b^2}\biggr),
\end{equation}
where $a = Q/(\sqrt{2} \sigma_\beta)$ and $b =
\sigma_\eta/\sigma_\beta$.  Setting 
\begin{equation}
Q = \sigma_\beta \sqrt{2} \, q, \qquad q = \erf^{-1}(2p-1),
\end{equation}
we obtain
\begin{equation} \label{eq:B}
B = \frac{\sigma_\eta}{p \sqrt{2\pi}} \,   \frac{1}{\sqrt{1 + (\sigma_\beta/\sigma_\eta)^2}} 
\exp\biggl(-\frac{q^2}{1+(\sigma_\eta/\sigma_\beta)^2}\biggr),
\end{equation}
from which
\begin{equation} \label{eq:beta_low_final}
\beta_{\textrm{low factor}}\approx -50\% \frac{\sigma_\eta}{p \sqrt{2\pi}} \,   \frac{1}{\sqrt{1 + (\sigma_\beta/\sigma_\eta)^2}} 
\exp\biggl(-\frac{q^2}{1+(\sigma_\eta/\sigma_\beta)^2}\biggr).
\end{equation}

From the data for the USA, we estimate the standard deviation of the
measured beta ($\sigma_{\beta}=0.43$), the volatility of the stock
index ($\sigma_I = 19.77\%$), the volatility of the low volatility
factor ($3.46\%$), and $\langle \sigma_i\rangle/\sigma_I = 1.53$.
Setting $\lambda_\beta = 1/90$, we obtain from
Eq. (\ref{eq:sigma_eta}) $\sigma_\eta = 1.53 \, \sqrt{1/90} = 0.1613$.
For $p = 0.3$ (bottom $30\%$), we obtain $q = -0.3708$ and, thus,
$\beta_{\textrm{low factor}} \approx 0.0334$ from
Eq. (\ref{eq:beta_low_final}).  Finally, we conclude that
$\rho_{\textrm{low factor}}=3.34\%\frac{19.77\%}{3.46\%}=19.1\%$.

\section{Construction of the beta-neutral factors}
\label{sec:Afactors}

We implement the four most popular strategies as four beta-neutral
factors that are constructed as follows.  First, we split all stocks
into six supersectors of similar sizes to minimize sectorial
correlations.  For each trading day, the stocks of the chosen
supersector are sorted according to the indicator (e.g., the
capitalization) available the day before (we use the publication date
and not the valuation date).  The related indicator-based factor is
formed by buying the first $pN$ stocks in the sorted list and shorting
the last $pN$ stocks, where $N$ is the number of stocks in the
considered supersector, and $p$ is a chosen quantile level.  As
described in Sec. \ref{sec:results}, we use $p = 0.15$ for short-term
reversal and long-term momentum factors and $p = 0.30$ for the
capitalization and low volatility factors.  The other stocks (with
intermediate indicator values) are not included (weighted by $0$).  To
reduce the specific risk, the weights of the selected stocks are set
inversely proportional to the stock's volatility $\sigma_i$, whereas
the weights of the remaining stocks are $0$.  Moreover, the inverse
stock volatility is limited to reduce the impact of extreme specific
risk.  Each trading day, we recompute the weight $w_{i}$ as follows
\begin{equation}
\label{eq:weights}
w_{i} = \left\{ \begin{array}{c l}  + \mu_+ \min\{1, \sigma_{\rm mean}/\sigma_i  \}, & 
\qquad \textrm{if $i$ belongs to the first $pN$ stocks in the sorted list}, \\
- \mu_- \min\{1, \sigma_{\rm mean}/\sigma_i \}, & 
\qquad  \textrm{if $i$ belongs to the last $pN$ stocks in the sorted list},\\
0, & \qquad \textrm{otherwise,} \\  \end{array}  \right.
\end{equation}
where $\sigma_{\rm mean} = \frac{1}{N}(\sigma_1 + \ldots + \sigma_N)$
is the mean estimated volatility over the cluster of sectors.  In this
manner, the weights of low-volatility stocks are reduced to avoid
strongly unbalanced portfolios concentrated in such stocks.  The two
common multipliers, $\mu_\pm$, are used to ensure the beta market
neutral condition:
\begin{equation}
\label{eq:beta_neutral}
\sum\limits_{i=1}^N \beta_{i} w_{i} = 0 ,
\end{equation}
where $\beta_{i}$ is the sensitivity of stock $i$ to the market
obtained either by an OLS or by our reactive method.  In every
case, the method to estimate beta uses the rolling daily returns and
only past information to avoid the look-ahead bias.  If the
aggregated sensitivity of the long part of the portfolio to the market
is higher than that of the short part of the portfolio, its weight is
reduced by the common multiplier $\mu_+ < \frac{1}{2p N}$, which is
obtained from Eq. (\ref{eq:beta_neutral}) by setting $\mu_- =
\frac{1}{2p N}$ (which implies that the sum of absolute weights
$|w_i|$ does not exceed $1$).  In the opposite situation (when the
short part of the portfolio has a higher aggregated beta), one sets
$\mu_+ = \frac{1}{2p N}$ and determines the reducing multiplier $\mu_-
< \frac{1}{2p N}$ from Eq. (\ref{eq:beta_neutral}).  The resulting
factor is obtained by aggregating the weights constructed for each
supersector.  We emphasize that the factors are constructed on a daily
basis, i.e., the weights are re-evaluated daily based on updated
indicators.  However, most indicators do not change frequently, so 
the transaction costs related to changing the factors are not
significant.

%%%%%%%%%%%%%%%%%%%%%%%%%%%%%%%%%%%%%

\subsection{Appendix C: Description of alternative methods}
\label{sec:methods}	
	
\subsubsection{Unconditional beta} 
	
\paragraph{The theory.}
	
\cite{Chan92} produce an empirical analysis that
describes various robust methods for estimating constant beta as they
provide an alternative to Ordinary Least Squares (OLS).  Their method
is built the work \cite{Koenker78} that provides robust alternatives
to the sample mean using more complex linear combination of order
statistics in order to face the case of non-Gaussian errors, which are
the source of outliers.  Instead of minimizing the sum of squared
residuals, they consider an estimator that is based on minimizing the
criterion including a penalty function $\varrho$ on the residuals
$\epsilon$:
\begin{equation}\label{min}
\sum_{t=1}^{T} \varrho_{\theta} (\epsilon_t)
\end{equation}
for $\varrho_{\theta} (\epsilon_t) = \theta \left| \epsilon_t \right|$
if $\epsilon_t \geq 0,$ or $(1 - \theta) \left| \epsilon_t \right|$ if
$\epsilon_t < 0$, where $0< \theta <1$.

\cite{Chan92} minimize the sum of absolute deviations of the residuals
$\epsilon_{it}$ from the market model, instead of the sum of squared
deviations.  The resulting minimum absolute deviations (MAD) estimator
of the regression parameters corresponds to the special case of
$\theta=1/2$ where half of the observations lie above the line, while
half lie below.  More generally, large or small values of the weight
$\theta$ attach a penalty to observations with large positive or
negative residuals. Varying $\theta$ between 0 and 1 yields a set of
regression quantile estimates $\hat{\beta(\theta)}$ that is analogous
to the quantiles of any sample of data.  However, they recognize that
MAD does not prove itself to be a clearly superior method and suggest
that it may be improved via linear combinations of sample quantiles
such like trimmed means.
	
For that reason, \cite{Chan92} test different combinations of
regressions quantiles serving as the basis for the robust estimators.
They discuss the general case of trimmed regression quantile (TRQ)
given as a weighted average of the regression quantile statistics:
\begin{equation}\label{trq}
\hat{\beta}_{\alpha} = (1-2\alpha)^{-1} \int_{\alpha}^{1-\alpha} \hat{\beta}(\theta) d\theta
\end{equation}
where $0<\alpha<1/2$ and $0<\theta<1$.

More specifically, \cite{Chan92} suggest a more straightforward and
equivalent method that considers estimators that are finite linear
combinations of regression quantiles (QR) and computationally simpler:
\begin{equation}\label{regq}
\beta_{\omega} = \sum_{i=1}^{N} \omega_i \hat{\beta}(\theta_i)
\end{equation}
where weights $0<\omega_i<1$, $i=1,...,N$ and $\sum_{i=1}^{N} \omega_i=1$.
The specific case of weighted average is given by the Tukey's trimean
(TRM) estimator:
\begin{align}
\hat{\beta}_{TRM}  =  0.25 \hat{\beta}(1/4) + 0.5 \hat{\beta}(1/2) + 0.25 \hat{\beta}(3/4) 
\end{align}

\paragraph{The application.}
	
Their analysis is based mainly on simulated returns data although they
add some tests with actual returns data.  The main advantages of a
simulation are that the true values of the underlying parameters are
known, and that the extent of departures from normality can be
controlled.  They begin with a baseline simulation with 25,000
replications using data generated from a normal distribution and they
also consider the case where the residual term is drawn from a
Student-distribution with three degrees of freedom in order to explain
the observed leptokurtosis in daily returns data.  We follow the same
methodology to assess the quality of the OLS, the MAD and the TRM beta
estimators using Gaussian and t-Student residuals in the seven types
of Monte Carlo simulations (MC1,...,MC7).
	
To replicate the exponential weight scheme of the reactive model
($\lambda_\beta=1/90$), Eq. (\ref{min}) is replaced by
\begin{equation}\label{min2}
\sum_{t=1}^{T} \left(1-\lambda_\beta\right)^{T-t} \varrho_{\theta} (\epsilon_t)
\end{equation}

\subsubsection{Conditional Beta} 
	
\paragraph{The theory.}
	
The first application of time-varying beta was proposed in
\cite{Bollerslev88} since the beta was computed as the ratio of the
conditional covariance to the conditional variance.
\cite{Engle02} generalizes \cite{Bollerslev90} constant correlation
model by making the conditional correlation matrix time-dependent with
the Dynamic Conditional Correlation (DCC) model that constrains the
time-varying conditional correlation matrix to be positive definite
and the number of parameters to grow linearly by a two step procedure.
The first step requires the GARCH variances to be estimated
univariately.  Their parameter estimates remains constant for the next
step.  The second stage is estimated conditioning on the parameters
estimated in the first stage.

%The most accomplished theory on
%time-varying beta comes from \cite{Engle16} who proposes the Dynamic
%Conditional Beta (DCB) as an approach to estimating regressions with
%time-varying parameters derived from the DCC model (DCC-GARCH beta).
Hereafter, we extend the modeling of the DCC beta for inclusion of an
asymmetric term in the conditional variance equation.  In the case of
asymmetry in the conditional variance, we select the GJR-GARCH(1,1)
specification by \cite{Glosten93}, which assumes a specific parametric
form with leverage effect in the conditional variance (DCC-GJR beta).
The basic idea is that negative shocks at period $(t-1)$ have a
stronger impact in the conditional variance at period $t$ than
positive shocks.  Notice that even though the conditional distribution
is Gaussian, the corresponding unconditional distribution can still
present excess kurtosis.

We select the ADCC model by \cite{Cappiello06} to incorporate
asymmetry in correlation. The case mixing asymetry in both located in
the variance equation (GJR-GARCH) and in the correlation equation
(ADCC) is examined (ADCC-GJR GARCH). In our paper the symmetric GARCH
DCC will be called simply DCC and the asymmetric ADCC-GJR will be
called simply ADCC

Let us consider $r_i$ and $r_I$ as the returns of a single stock and
the stock index, respectively.  We assume that their respective
conditional variances follow a (GJR-)GARCH(1,1) specification.  The
stock return $r_i$ is defined by its conditional volatility,
$\sigma_i$, and a zero-mean white noise $\xi_i(t)$:
\begin{equation}
r_i(t)=\sigma_i(t-1) \xi_i(t)
\end{equation}
The conditional variation specification of the stock return is the
following:
\begin{equation}  \label{eq:beta_aux1}
\sigma_i^{2}(t)= (1-a-b-\gamma/2) \tilde\sigma_i^{2} + a \sigma_i^{2}(t-1) [\xi_i(t)]^2+b 
\sigma_i^{2}(t-1)+\gamma \sigma_i^{2} [\xi_i^-(t)]^2
\end{equation}
where $\tilde\sigma_i$ is the unconditional volatility, and $a$, $b$,
and $\gamma$ are parameters reflecting respectively, the ARCH, GARCH
and asymmetry effects.  When $\gamma=0$, the specification collapses
to a GARCH model, otherwise, it stands for the GJR-GARCH model, where
the asymmetric term is defined such as $\xi_i^-(t)=\xi_i(t)$ if
$\xi_i(t)>0$, otherwise $\xi_i^-(t)=0$.
		
The stock index return $r_I$ is defined by its conditional volatility,
$\sigma_I$, and a zero-mean white noise $\xi_I(t)$ that is correlated
to $\xi_i(t)$:
\begin{equation}
r_I(t)=\sigma_I(t-1) \xi_I(t)
\end{equation}

The conditional variance specification of the stock index return is
the following:
\begin{equation}  \label{eq:beta_aux2}
\sigma_I^{2}(t)= (1-a-b-\gamma/2) \tilde\sigma_I^{2} + a \sigma_I^{2}(t-1) [\xi_I(t)]^2+b \sigma_I^{2}(t-1) +\gamma \sigma_I^{2} [\xi_I^-(t)]^2
\end{equation}
	
We define the normalized conditional variance diagonal terms such as:
%	
%\begin{equation}
%q_{ii}(t) = \sigma_i^{2}(t)/\tilde\sigma_i^{2} 
%\end{equation}
%	
%\begin{equation}
%q_{II}(t) = \sigma_I^{2}(t)/\tilde\sigma_I^{2} 
%\end{equation}

\begin{equation}
q_{ii}(t)=(1-a_\rho-b_\rho- \gamma_\rho/2)  + a_\rho \xi_i(t-1) \xi_i(t-1) + b_\rho q_{ii}(t-1) +\gamma_\rho \xi_i^-(t-1) \xi_i^-(t-1)
\end{equation}

\begin{equation}
q_{II}(t)=(1-a_\rho-b_\rho- \gamma_\rho/2) + a_\rho \xi_I(t-1) \xi_I(t-1) + b_\rho q_{II}(t-1) +\gamma_\rho \xi_I^-(t-1) \xi_I^-(t-1)
\end{equation}
		
The normalized conditional covariance term $q_{iI}(t)$ is given by:
\begin{equation}
q_{iI}(t)=(1-a_\rho-b_\rho- \gamma_\rho/4) \tilde \rho + a_\rho \xi_i(t-1) \xi_I(t-1) + b_\rho q_{iI}(t-1) +\gamma_\rho \xi_i^-(t-1) \xi_I^-(t-1)
\end{equation}
	
When $\gamma_\rho=0$, the specification collapses to a DCC model,
otherwise it stands for the ADCC model, where the asymmetric term is
defined such as $\xi_i^-(t)=\xi_i(t)$ if $\xi_i(t)>0$, otherwise
$\xi_i^-(t)=0$.
	
The conditional correlation between $\xi_I(t+1)$ and $\xi_i(t+1)$ is
then updated by:
\begin{equation}
\rho_{iI}(t)=q_{iI}(t)/\sqrt{ q_{II}(t) q_{ii}(t)}
\end{equation}
	
The beta DCC and beta ADCC estimation are defined in the same way:
\begin{equation}
\beta_{DCC}(t)=\rho_{iI}(t) \sigma_i(t)/\sigma_I(t)
\end{equation}
	
	%	L=L+wi*real(-u1^2-u2^2-log(abs(sigmai))-log(abs(sigmam)));
	%	L=L+wi*real(-log((det(rho)))-transpose(u)*inv(rho)*u+transpose(u)*u);
	
The log-likelihood function is optimized to calibrate the parameters
$\tilde \rho$, $\tilde\sigma_I$ and $\tilde\sigma_i$ for estimation:
\begin{equation} \label{eq:Lbis}
L_{DCC} = \frac{1}{2}\sum_{t}^T  \left( L_{V}(t)+L_{C}(t)\right) 
\end{equation}
\begin{equation}
L_{V}(t)=-  2 \log(2\pi) -\xi_I(t)^2-\xi_i(t)^2- 2\log(\sigma_I(t))-2\log(\sigma_i(t))
\end{equation}
\begin{equation}
L_{C}(t)= -\log(det(R(t)))- U'(t)R(t)^{-1}U(t)- U'(t)U(t) 
\end{equation}
with $det$ as the determinant of a matrix, and
\begin{equation}
R(t)=\begin{bmatrix}
	1       & \rho_{iI}(t)  \\
	\rho_{iI}(t)       & 1 	
	\end{bmatrix},  \qquad 
U(t)=\begin{bmatrix}
	\xi_i(t)  \\
	\xi_I(t)    	
	\end{bmatrix}
\end{equation}

\paragraph{The application.}
	
For Monte Carlo simulation purposes:
	
\begin{itemize}
\item 
$\xi_i(t)$ is either generated randomly in MC6 and MC7 according to a
standard Gaussian or measured through returns $r_i(t)$ and
$\sigma_i(t-1)$ for beta DCC estimation.
		
\item 
$\gamma=0$ for MC6 and beta DCC estimation but $\gamma>0$ for MC7 and
beta ADCC that captures the asymmetry term of the GJR-GARCH. 
		
\item 
$\xi_I(t)$ is either generated randomly in MC6 and MC7 according to a
standard Gaussian random variable that is correlated to the random
variable $\xi_i(t)$ (correlation between $\xi_i(t)$ and $\xi_I(t)$ is
$\rho_{iI}(t-1)$) or measured through returns $r_I(t)$ and
$\sigma_I(t-1)$ for beta DCC estimation.
		
\item 
$\gamma_\rho=0$ for MC6 and beta DCC but $\gamma_\rho>0$ for MC7 and
beta ADCC that captures the asymmetry term of the ADCC.
\end{itemize}
	
The fixed parameters that are supposed to be known when testing the
beta DCC are set to US market estimates by from \cite{Sheppard17}:
\begin{itemize}
\item 
fixed parameters for  univariate symmetric GARCH(1,1) process (MC6, i.e. DCC):
\subitem $b=0.89$, $b$ is the decay coefficient and $1/(1-b)$ is related to the number of days the process needs to mean revert;
\subitem  $a=0.099$ would describe the level of the volatility of the volatility.

\item fixed parameters for  univariate asymmetric GJR-GARCH(1,1,1) process (MC7, i.e., ADCC):
\subitem $b=0.901$, $b$ is the decay coefficient and  $1/(1-b)$ is related to the number of days the process needs to mean revert;
\subitem $a=0.0$, $a+\gamma/2$ describe the level of the volatility of the volatility;
\subitem $\gamma=0.171$, $\gamma$  would describe the asymmetry.
\end{itemize}
	
The fixed parameters that are supposed to be known when testing the
beta DCC and betas ADCC are set to US market estimates from
\cite{Cappiello06}:
	
\begin{itemize}
\item fixed parameters for the symmetric cross term process (MC6, i.e., DCC):
\subitem $b_\rho=0.9261$, $b_\rho$ is the decay coefficient and is linked to the relaxation time;
\subitem $a_\rho=0.0079$ would describe the level of the volatility.

\item fixed parameters for the asymmetric cross term process (MC7, i.e., ADCC):
\subitem $b_\rho=0.9512$, $b_\rho$ is the decay coefficient and is linked to the relaxation time;
\subitem $a_\rho=0.0020$, $a_\rho+\gamma_\rho/4$   would describe the level of  the volatility of the correlation;
\subitem $\gamma_\rho=0.0040$, $\gamma_\rho$ would describe the asymmetry.
		
\end{itemize}
	
The fixed parameters that are not known when testing the beta DCC and
estimated through optimization of log-likelihood are set by MC
simulation to:
\begin{itemize}
\item $\tilde \rho=0.15/0.4$, unconditional correlation;
\item $\tilde\sigma_I=0.15/\sqrt{255}$,  $\tilde\sigma_i=0.4/\sqrt{255}$ unconditional stock index volatility;
\item $\tilde\sigma_i=0.4/\sqrt{255}$ unconditional single stock  volatility.
\end{itemize}

To replicate the exponential weight scheme in the reactive model
($\lambda_\beta=1/90$), Eq. (\ref{eq:Lbis}) is replaced by
\begin{equation} 
L_{DCC} = \frac{1}{2}\sum_{t}^T \left(1-\lambda_\beta\right)^{T-t}  \left( L_{V}(t)+L_{C}(t)\right) 
\end{equation}


\newpage



\footnotesize
\begin{thebibliography}{99}

\bibitem[Acharyaa and Pedersen(2005)]{Acharyaa05}    
			Acharyaa, V., and L. H., Pedersen. 
			``Asset Pricing with Liquidity Risk.''
			Journal of Financial Economics, 77 (2005), pp. 375-410.

\bibitem[Agarwal and Naik(2004)]{Agarwal04} 
			Agarwal, V., and N. Y. Naik. 
			``Risks and Portfolio Decisions Involving Hedge Funds.''
			Review of Financial Studies, 17 (2004), pp. 63-98.

\bibitem[Amihud(2002)]{Amihud02}      
			Amihud, Y. 
			``Illiquidity and Stock Returns: Cross-section and Time-series Effects.''
			Journal of Financial Markets, 5 (2002), pp. 31-56.


\bibitem[Ang and Liu (2004)]{Ang04} 	
			Ang, A., and J. Liu.
			``How to Discount Cashflows with Time-Varying Expected Returns.'' 
			Journal of Finance, 59, 6 (2004), pp. 2745-2783.


\bibitem[Ang and Chen(2007)]{Ang07}         
			Ang, A., and J. Chen.
			``CAPM over the Long Run: 1926-2001.''
			Journal of Empirical Finance, 14 (2007), pp. 1-40.

\bibitem[Ang\etal(2006)]{Ang06}         
			Ang, A., R. Hodrick, Y. Xing, and X. Zhang. 
			``The Cross-Section of Volatility and Expected Returns.'' 
			Journal of Finance, 61 (2006), pp. 259-299.

\bibitem[Ang\etal(2009)]{Ang09} 	
			Ang, A., R., Hodrick, Y. Xing, and X. Zhang. 
			``High Idiosyncratic Volatility and Low Returns: International and Further U.S. Evidence.''
			Journal of Financial Economics, 91 (2009), pp. 1-23.

\bibitem[Ang\etal(2013)]{Ang13}         
			Ang, A., A. Shtauber, and P. C. Tetlock. 
			``Asset Pricing in the Dark: The Cross-Section of OTC Stocks.''
			Review of Financial Studies, 26 (2013), pp. 2985-3028.

\bibitem[Asness\etal(2001)]{Asness01} 
			Asness, C., R. Krail, and J. Liew. 
			``Do Hedge Funds Hedge?'' 
			Journal of Portfolio Management, 28 (2001), pp. 6-19.

\bibitem[Baker\etal(2013)]{Baker13} 	
			Baker, M., B. Bradley, and R. Taliaferro. 
			``The Low Beta Anomaly: A Decomposition into Micro and Macro Effects.''
			Working paper, Harvard Business School, 2013.
			
\bibitem[Bali\etal(2017)]{Bali17}	
			Bali, T. G. and R. F. Engle, and Y. Tang.  
			``Dynamic Conditional Beta Is Alive and Well in the Cross Section of Daily Stock Returns.'' 
			Management Science, 63, 11 (2017), pp. 3760-3779.
			

\bibitem[Banz(1981)]{Banz81} 	
			Banz, R. W.
			``The Relationship Between Return and Market Value of Common Stocks.''
			Journal of Financial Economics, 9 (1981), pp. 3-18.

\bibitem[Bekaert and Wu(2000)]{Bekaert00} 	
			Bekaert, G., and G. Wu. 
			``Asymmetric Volatility and Risk in Equity Markets.''
			Review of Financial Studies, 13 (2000), pp. 1-42.

\bibitem[Black(1972)]{Black72} 	
			Black, F. 
			``Capital Market Equilibrium with Restricted Borrowing.''
			Journal of Business, 4 (1972), pp. 444-455.

\bibitem[Black\etal(1972)]{Black72b} 	
			Black, F., M. Jensen, and M. Scholes.
			``The Capital Asset Pricing Model: Some Empirical Tests.''
			In Studies in the Theory of Capital Markets, edited by Michael Jensen. New York: Praeger, 1972.

\bibitem[Black(1976)]{Black76} 	
			Black, F. 
			``Studies in Stock Price Volatility Changes.''
			American Statistical Association, Proceedings of the Business and Economic Statistics Section, 177-181, 1976.

\bibitem[Blume(1971)]{Blume71} 
			Blume, M. E. 
			``On the Assessment of Risk.'' 
			Journal of Finance, 26 (1971), pp. 1-10.




\bibitem[Bollerslev\etal(1988)]{Bollerslev88} 	
			Bollerslev, T., R. Engle, and J. Wooldridge. 
			``A Capital Asset Pricing Model with Time-Varying Covariances.''
			Journal of Political Economy, 96 (1988), pp. 116-131.


\bibitem[Bollerslev(1990)]{Bollerslev90}
			Bollerslev, T.
			``Modelling the coherence in short-run nominal exchange rates: A multivariate generalized ARCH model.'' 
			Review of Economics and Statistics, 72 (1990), pp. 498-505.


\bibitem[Bouchaud\etal(2001)]{Bouchaud01} 	
			Bouchaud, J.-P., A. Matacz, and M. Potters. 
			``Leverage Effect in Financial Markets: The Retarded Volatility Model.''
			Physical Review Letters, 87 (2001), pp. 1-4.

\bibitem[Bussi\`ere\etal(2015)]{Bussiere15} 
			Bussi\`ere, M., M. Hoerova, and B. Klaus. 
			``Commonality in Hedge Fund Returns: Driving Factors and Implications.'' 
			Journal of Banking \& Finance, 54 (2015), pp. 266-280.

\bibitem[Campbell and Hentchel(1992)]{Campbell92} 	
			Campbell, J. Y., and L. Hentchel, 
			``No News is Good News: An Asymmetric Model of Changing Volatility in Stock Returns.''
			Journal of Financial Economics, 31 (1992), pp. 281-318.


\bibitem[Cappiello\etal(2006)]{Cappiello06}	
			Cappiello, L., R. Engle, and K. Sheppard.
			``Asymmetric Dynamics in the Correlations of Global Equity and Bond Returns.''
			Journal of Financial Econometrics, 4 (2006), pp. 537-572.


\bibitem[Carhart(1997)]{Carhart97} 	
			Carhart, M. 
			``On Persistence in Mutual Fund Performance.''
			Journal of Finance, 52 (1997), pp. 57-82.

%\bibitem[Cederburg\etal(2015)]{Cederburg15} 	
%			Cederburg, S., P. Davies, and M. O'Doherty. 
%			``Asset-Pricing Anomalies at the Firm Level.''
%			Journal of Econometrics, 186 (2015), pp. 113-128.

\bibitem[Chan and Lakonishok(1992)]{Chan92} 	
			Chan, L., and J. Lakonishok. 
			``Robust Measurement of Beta Risk.''
			Journal of Financial and Quantitative Analysis, 27 (1992), pp. 265-282.

\bibitem[Chen\etal(2005)]{Chen05} 	
			Chen, A-S., T.-W. Zhang, and H. Wu. 
			``The Beta Regime Change Risk Premium.''
			Working paper, National Chung Cheng University, 2005.

\bibitem[Christie(1982)]{Christie82} 	
			Christie, A.  
			``The Stochastic Behavior of Common Stock Variances - Value, Leverage, and Interest Rate Effects.''
			Journal of Financial Economic Theory, 10 (1982), pp. 407-432.

\bibitem[DeJong and Collins(1985)]{DeJong85} 	
			DeJong, D., and D. W. Collins. 
			``Explanations for the Instability of Equity Beta: Risk-Free Rate Changes and Leverage Effects.''
			Journal of Financial and Quantitative Analysis, 20 (1985), 73-94.

\bibitem[Fabozzi and Francis(1978)]{Fabozzi78} 	
			Fabozzi, J. F., and C. J. Francis. 
			``Beta Random Coefficient.''
			Journal of Financial and Quantitative Analysis, 13 (1978), pp. 101-116.
			
		
\bibitem[Engle(2002)]{Engle02}
			Engle, R. F. 
			''Dynamic Conditional Correlation: A Simple Class of Multivariate Generalized Autoregressive 
			Conditional Heteroskedasticity Models.'' 
			Journal of Business and Economic Statistics, 20 (2002), pp. 339-350.			
			
\bibitem[Engle(2016)]{Engle16}
			Engle, R. F. 
			``Dynamic Conditional Beta.'' 
			Journal of Financial Econometrics, 14, Issue 4 (2016), pp. 643-667.

\bibitem[Fama and French(1992)]{Fama92} 	
			Fama, E. F., and K. R. French. 
			``The cross-section of expected returns.''
			Journal of Finance, 47 (1992), pp. 427-465.

\bibitem[Fama and French(1993)]{Fama93}        
			Fama, E. F., and K. R. French.
			``Common risk factors in the returns on stocks and bonds.''
                        Journal of Financial Economics, 33 (1993), pp. 3-56.

\bibitem[Fama and French(1997)]{Fama97} 
			Fama, E. F., and K. R. French. 
			``Industry Costs of Equity.'', 
			Journal of Financial Economics, 43 (1997), pp. 153-193.

\bibitem[Fama and French(2008)]{Fama08} 	
			Fama, E. F., and K. R. French. 
			``Dissecting Anomalies.''
			Journal of Finance, 63 (2008), pp. 1653-1678.

\bibitem[Fama and French(2012)]{Fama12} 	
			Fama, E. F., and K. R. French. 
			``Size, Value, and Momentum in International Stock Returns.''
			Journal of Financial Economics, 105 (2012), pp. 457-472.

%\bibitem[Ferson and Foerster(1994)]{Ferson94} 
%			Ferson, W. E. and S.R. Foerster. 
%			``Finite Sample Properties of the Generalized Method of Moments in Tests of Conditional Asset Pricing Models.''  
%			Journal of Financial Economics, 36 (1994), pp. 29-55.

\bibitem[Ferson and Harvey(1999)]{Ferson99} 	
			Ferson, W. E., and C. R. Harvey. 
			``Conditioning Variables and the Cross-Section of Stock Returns.''
			Journal of Finance, 54 (1999), pp. 1325-1360.

\bibitem[Francis(1979)]{Francis79} 	
			Francis, J. C. 
			``Statistical Analysis of Risk Surrogates for NYSE Stocks.''
			Journal of Financial and Quantitative Analysis, 14 (1979), pp. 981-997.

\bibitem[Frazzini and Pedersen(2014)]{Frazzini14} 	
			Frazzini, A., and L. Pedersen. 
			``Betting Against Beta.''
			Journal of Financial Economics, 111 (2014), pp. 1-25.

\bibitem[Fung and Hsieh(1997)]{Fung97} 
			Fung, W., and D. A. Hsieh. 
			``Empirical Characteristics of Dynamic Trading Strategies: The Case of Hedge Funds.'' 
			The Review of Financial Studies, 10 (1997), pp. 275-302.

\bibitem[Galai and Masulis(1976)]{Galai76}	
			Galai, D. and R. W. Masulis. 
			``The Option Pricing Model and the Risk Factor of Stock.''
			Journal of Financial Economics, 3 (1976), 53-81.


\bibitem[Glosten\etal(1993)]{Glosten93}	
			Glosten, L. R., R. Jagannathan, and D. E. Runkle. 
			``On The Relation between The Expected Value and The Volatility of Nominal Excess Return on stocks.''
			Journal of Finance, 48 (1993), pp. 1779-1801.


\bibitem[Goyal and Santa-Clara(2003)]{Goyal03} 	
			Goyal, A., and P. Santa-Clara, 
			``Idiosyncratic Risk Matters!''
			Journal of Finance, 58 (2003), pp. 975-1007.

\bibitem[Grinblatt and Moskowitz(2004)]{Grinblatt04} 	
			Grinblatt, M., and T., Moskowitz. 
			``Predicting Stock Price Movements from Past Returns: The Role of Consistency and Tax-Loss Selling.''
			Journal of Financial Economics, 71 (2004), pp. 541-579.

\bibitem[Haugen and Heins(1975)]{Haugen75} 	
			Haugen, R. A., and A. Heins. 
			``Risk and the Rate of Return on Financial Assets: Some Old Wine in New Bottles.''
			Journal of Financial and Quantitative Analysis, 10 (1975), pp. 775-784.

\bibitem[Haugen and Baker(1991)]{Haugen91} 
			Haugen, R. A., and N. L. Baker.  
			``The Efficient Market Inefficiency of Capitalization-Weighted Stock Portfolios.''
			Journal of Portfolio Management, 17 (1991), pp. 35-40.

\bibitem[Hong and Sraer(2016)]{Hong16} 	
			Hong, H., and D. Sraer. 
			``Speculative Betas.''
			Forthcoming, Journal of Finance, (2016).

\bibitem[Jagannathan and Wang(1996)]{Jagannathan96} 
			Jagannathan, R., and Z. Wang, 
			``The Conditional CAPM and the Cross-Section of Expected Returns.''
			Journal of Finance, 51 (1996), pp. 3-53.
			
\bibitem[Jegadeesh(1990)]{Jegadeesh90}
			Jegadeesh, N. 
			``Evidence of Predictable Behavior of Securities Returns.''
			Journal of Finance, 45 (1990), pp. 881-898.
			
\bibitem[Jegadeesh and Titman(1993)]{Jegadeesh93}
			Jegadeesh, N., and S. Titman. 
			``Returns to buying winners and selling losers: Implications for stock market efficiency.''
			Journal of Finance, 48 (1993), pp. 65-91.


\bibitem[Koenker(1978)]{Koenker78}	
			Koenker, R., and G. Bassett. 
			``Regression Quantiles.''
			Econometrica, 46 (1978), pp. 33-50.

\bibitem[Koenker(1982)]{Koenker82}
			Koenker, R.  
			``Robust Methods in Econometrics.''
			Econometric Reviews, 1 (1982), pp. 213-255.



\bibitem[Lettau and Ludvigson(2001)]{Lettau01} 	
			Lettau, M., and S. Ludvigson. 
			``Resurrecting the (C)CAPM: a Cross-Sectional Test when Risk Premia are Time-Varying.''
			Journal of Political Economy, 109 (2001), pp. 1238-1287.

\bibitem[Lewellen and Nagel(2006)]{Lewellen06} 
			Lewellen, J., and S. Nagel. 
			``The Conditional CAPM does not Explain Asset Pricing Anomalies.'' 
			Journal of Financial Economics,  82 (2006), pp. 289-314.

\bibitem[Malkiel and Xu(1997)]{Malkiel97} 	
			Malkiel, B. G., and Y. Xu. 
			``Risk and Return Revisited.''
			Journal of Portfolio Management, 23 (1997), pp. 9-14.

\bibitem[Meng\etal(2011)]{Meng11} 	
			Meng, J. G., G. Hu, and J. A. Bai. 
			``A Simple Method for Estimating Betas When Factors Are Measured with Error.''
			Journal of Financial Research, 34 (2011), pp. 27-60.

%\bibitem[McCulloch and Rossi(1991)]{McCulloch91} 	
%			McCulloch, R., and P. E. Rossi. 
%			``A Bayesian Approach to Testing the Arbitrage Pricing Theory.''
%			Journal of Econometrics, 49 (1991), pp. 141-168.

\bibitem[Mitchell and Pulvino(2001)]{Mitchell01} 
			Mitchell, M., and T. Pulvino. 
			``Characteristics of Risk and Return in Risk Arbitrage.'' 
			Journal of Finance, 56 (2001), pp. 2135-2175.

\bibitem[Patton(2009)]{Patton09} 
			Patton, A. J. 
			``Are `Market Neutral' Hedge Funds Really Market Neutral?'' 
			Review of Financial Studies, 22 (2009), pp. 2295-2330.

\bibitem[Reinganum(1981)]{Reinganum81} 	
			Reinganum, R. 
			``Misspecification of Capital Asset Pricing: Empirical Anomalies based on Earnings Yields and Market Values.''
			Journal of Financial Economics, 9 (1981), pp. 19-46. 

\bibitem[Roll(1977)]{Roll77} 
			Roll, R. 
			``A Critique of the Asset Pricing Theory's Tests Part I: On Past and Potential Testability of the Theory.'' 
			Journal of Financial Economics, 4 (1977), pp. 129-176.

\bibitem[Sharpe(1964)]{Sharpe64} 	
			Sharpe, W. F. 
			``Capital Asset Prices: A Theory of Market Equilibrium under Risk.''
			Journal of Finance, 19 (1964), pp. 425-442.

\bibitem[Shanken(1992)]{Shanken92} 	
			Shanken, J. 
			``On the Estimation of Beta-Pricing Models.''
			Review of Financial Studies, 5 (1992), 1-33.

\bibitem[Sheppard(2017)]{Sheppard17}	
			Sheppard, K. 
			``Univariate volatility modeling.''
			Notes, Chapter 7, University of Oxford (2017).

%\bibitem[Tofallis(2008)]{Tofallis08} 	
	%		Tofallis, C.  
	% 	``Investment Volatility: A Critique of Standard Beta Estimation and a Simple Way Forward.''
	%		European Journal of Operational Research, 187 (2008), pp. 1358-1367.

\bibitem[Valeyre\etal(2013)]{Valeyre13} 	
			Valeyre, S., D. S. Grebenkov, S. Aboura, and Q. Liu. 
			``The Reactive Volatility Model.''
			Quantitative Finance, 13, (2013), pp. 1697-1706.

\end{thebibliography}
\end{document}